\documentclass[lettersize,journal]{IEEEtran}
\usepackage[switch]{lineno} % line number
\usepackage{amsmath,amssymb,amsfonts,amsthm}
\usepackage{array}
\usepackage[caption=false]{subfig}
\usepackage{textcomp}
\usepackage{stfloats}
\usepackage{url}
\usepackage{verbatim}
\usepackage{graphicx}
\usepackage{cite}
\usepackage{xcolor}
\usepackage{multirow}
\usepackage{setspace}
\def\BibTeX{{\rm B\kern-.05em{\sc i\kern-.025em b}\kern-.08em
    T\kern-.1667em\lower.7ex\hbox{E}\kern-.125emX}}
\usepackage{ragged2e}
\usepackage{enumitem}

\makeatletter
\newif\if@restonecol
\makeatother

\usepackage[linesnumbered,ruled,vlined]{algorithm2e}

\SetCommentSty{mycommfont}

\usepackage{algpseudocode}
\usepackage{amsmath}
\SetKwComment{Comment}{/* }{ */}

\DeclareMathOperator{\Var}{\widehat{Var}}

\begin{document}

% \linenumbers

\title{CSI-RFF: Leveraging Micro-Signals on CSI for RF Fingerprinting of Commodity WiFi
}
\author{Ruiqi~Kong and
        He~(Henry)~Chen
\thanks{The authors are with the Department of Information Engineering, The Chinese University of Hong Kong, Hong Kong SAR, China. H. Chen is also with Shun Hing Institute of Advanced Engineering, The Chinese University of Hong Kong, Hong Kong SAR, China. (E-mail: \{kr020, he.chen\}@ie.cuhk.edu.hk)}
\thanks{{Part of the work has been presented in 
2023 IEEE 24th International Workshop on Signal Processing Advances in Wireless Communications (SPAWC 2023) \cite{kong2023physical}.} {The authors would like to thank Soung Chang Liew for his insightful discussions on LS-based sparse channel estimation.}}
\thanks{This research was supported in part by project \#MMT 79/22 of Shun Hing Institute of Advanced Engineering, The Chinese University of Hong Kong, and a research donation from Huawei.}
}

\maketitle

\begin{abstract}
This paper introduces CSI-RFF, a new framework that leverages micro-signals embedded within  \underline{C}hannel \underline{S}tate \underline{I}nformation (CSI) curves to realize \underline{R}adio-\underline{F}requency \underline{F}ingerprinting of commodity off-the-shelf (COTS) WiFi devices for open-set authentication. The micro-signals that serve as RF fingerprints are termed ``micro-CSI''. Through experimentation, we have found that the presence of micro-CSI can primarily be attributed to imperfections in the RF circuitry. Furthermore, this characteristic signal is detectable in WiFi 4/5/6 network interface cards (NICs). 
We have conducted further experiments to determine the most effective CSI collection configurations to stabilize micro-CSI. Yet, extracting micro-CSI for authentication purposes poses a significant challenge. This complexity arises from the fact that CSI measurements inherently include both micro-CSI and the distortions introduced by wireless channels. These two elements are intricately intertwined, making their separation non-trivial. To tackle this challenge, we have developed a signal space-based extraction technique for line-of-sight (LoS) scenarios, which can effectively separate the distortions caused by wireless channels and micro-CSI. Over the course of our comprehensive CSI data collection period extending beyond one year, we found that the extracted micro-CSI displays unique characteristics specific to each WiFi device and remains invariant over time. This establishes micro-CSI as a suitable candidate for device fingerprinting. Finally, we conduct a case study focusing on area access control for mobile robots. In particular, we applied our CSI-RFF framework to identify mobile robots operating in real-world indoor LoS environments based on their transmitted WiFi signals. To accomplish this, we have compared and employed anomaly detection algorithms for the authentication of 15 COTS WiFi 4/5/6 NICs that were carried by a mobile robot under both static and mobile conditions, maintaining an average signal-to-noise ratio (SNR) of 34 dB. Our experimental results demonstrate that the micro-CSI-based authentication algorithm can achieve an average attack detection rate close to $99\%$ with a false alarm rate of 0\% in both static and mobile conditions when using 20 CSI measurements to construct one fingerprint\footnote{The collected real-world dataset and the code are publicly available at \url{https://github.com/Oriseven/CSI-RFF}.}.
\end{abstract}


\section{Introduction}
\IEEEPARstart{W}{iFi} networks are nowadays ubiquitous and have been widely deployed in homes and enterprises and at public hotspots. Thanks to its high data rate, low cost, and support of mobility, WiFi has been adopted as the communication technology in more and more security-critical applications, e.g., industrial automation \cite{wifi}. In these applications, once attackers gain access to networks, malicious control commands may lead to severe equipment damage or even loss of human lives \cite{panPhysicalLayerSecurityIndustrial2018,yaacoub2022robotics}. 
%Due to the rise of automation in production facilities, Autonomous Mobile Robots (AMRs) are now being utilized to enhance efficiency, productivity, and safety in manufacturing. These self-contained mobile robots are capable of navigating and functioning autonomously in industrial environments. In such environments, certain areas are often restricted due to safety or security concerns. As a result, it is imperative to authenticate mobile robots for area access control to ensure the security of these restricted areas \cite{gavrilova2010state, securingamb}. 
Wireless authentication is currently achieved by applying cryptographic methods above the physical layer, such as the WPA, WPA2 (802.11i) \cite{802.11i}, and 802.11w \cite{802.11w}. However, these authentication schemes are not impervious to potential vulnerabilities, which may be exploited through various types of attacks, such as rogue access points and MAC address spoofing. These potential vulnerabilities underscore the need for the implementation of additional layers of protection \cite{7467419, Desmond2008, Penetrating2021}. Moreover, the authentication server has no direct way to find out whether the authenticity has been compromised or not by solely relying on cryptographic methods. 
As such, other authentication mechanisms for security-critical applications, complementary to today’s cryptographic approaches, are needed to further boost the security of WiFi networks.

Physical layer authentication based on radio-frequency (RF) fingerprinting has been introduced as a supplement to traditional cryptography-based authentication, with the aim of enhancing the security of wireless networks \cite{angueiraSurveyPhysicalLayer2022}. RF fingerprinting techniques offer considerable advantages, as they exploit the inherent and unique characteristics of imperfections in the RF circuitry of wireless transmitters. These imperfections are difficult to compromise or impersonate and are easily maintained \cite{zhang}. Specifically, these circuitry imperfections are embedded in all signals emitted and manifest as distortions that deviate from standard signals. Although these signal distortions have a negligible impact on the decoding of transmitted information, they are sufficiently distinctive across different devices to enable reliable device authentication. Moreover, by leveraging the inherent properties of RF circuitry, RF fingerprinting requires fewer computational resources compared with the demands of cryptography-based mechanisms for key generation and management \cite{9450821}.

Thanks to its untamable properties, RF fingerprinting has attracted enormous interest and demonstrated satisfactory performance in the context of wireless device identification and/or authentication in recent years. Nevertheless, existing RF fingerprinting mechanisms often need to rely on physical-layer signal samples to extract fingerprints. In practice, wireless chipsets do not report physical-layer samples to higher layers for authentication purposes. As such, previous efforts used dedicated and expensive instruments, such as vector signal analyzers and software-defined radios, to acquire physical-layer samples. This largely hinders the practical usability of these radio sample-based RF fingerprinting mechanisms in commodity off-the-shelf (COTS) systems.

In the meantime, a variety of channel state information (CSI) tools have been developed and have gained widespread use in wireless sensing and localization applications in recent years \cite{atheros,linux,nexmon,picoscenes}. This development implies that COTS WiFi devices are capable of reporting CSI to higher network layers once a suitable CSI tool has been appropriately installed. We note that in wireless communication systems, CSI needs to be estimated at the receiver side to equalize the impacts of wireless channels before decoding the transmitted data symbols. Therefore, the acquisition of CSI does not need extra resources and/or equipment, since it has been available at the physical layer of wireless communication systems. Furthermore, since the transmitted signals go through RF circuits before being emitted to wireless channels, the CSI estimated at the receiver side incorporates the distortions induced by both wireless channels and the transmitter's RF circuitry imperfections.
The above two observations motivate us to explore the feasibility of extracting RF fingerprints from CSI measurements. Such CSI-based RF fingerprinting, if made possible, will be more lightweight and ready-to-implement than the existing signal sample-based methods.

Realizing CSI-based RF fingerprinting poses two primary challenges. The first involves the complexity of CSI measurements, where distortions caused by RF circuitry imperfections become entangled with varying wireless channel distortions. Consequently, accurately extracting RF fingerprints from CSI measurements is a non-trivial task. The second challenge lies in stabilizing the extracted fingerprints to achieve accurate authentication. On one side, the RF circuitry of receiver chains can introduce additional distortions to the signals from the transmitter, complicating the recovery of the transmitter's fingerprints. On the other side, RF distortions present in the CSI are often subtle and may be compromised by noise or errors during signal processing.
%the distortions caused by RF circuitry imperfections may be unstable to environmental changes (e.g., temperature).

%\subsection{Contributions}
Our objective is to identify a robust RF fingerprint extractable from CSI measurements, allowing for precise authentication of COTS WiFi devices. To achieve this, we introduce CSI-RFF, a new physical-layer authentication framework. CSI-RFF capitalizes on the micro-signals embedded in CSI curves (relative to the subcarrier index), employing them as fingerprints for the authentication of commercial WiFi devices. The main contributions of this work are three-fold: 
\begin{itemize}[leftmargin=*]
	\item We introduce a new CSI-based RF fingerprint, referred to as the micro-CSI-based fingerprint. This unique identifier is manifested as micro-signals observed on the CSI curves of COTS WiFi devices. The discovery of the micro-CSI was an incidental outcome while collecting CSI measurements using the Atheros CSI Tool \cite{atheros}. Upon initial observation, micro-CSI seemed random, particularly when analyzing the CSI curves of a single measurement. Uncertainty arose regarding the cause of micro-CSI, with potential sources ranging from noise to possible imperfections within the CSI tool itself. To further investigate the origins of micro-CSI, we conducted a series of experiments. These tests suggested that micro-CSI is most likely attributed to imperfections in the RF circuitry, thereby implying its potential application as an RF fingerprint for device authentication. Notably, our experiments revealed the presence of micro-CSI in all tested WiFi 4/5/6 NICs. We further examined factors influencing the stability of micro-CSI and determined the optimal system configuration for acquiring CSI measurements to maximize micro-CSI stability.
\item To extract micro-CSI from CSI measurements, we put forth an approach that leverages the signal space properties of wireless channels and hardware distortions to effectively disentangle small-scale micro-CSI from wireless channels under strong LoS conditions. Additionally, our algorithm is designed to be robust to the channel leakage effect. Due to the small scale of micro-CSI, we need to use a group of CSI measurements ($N_{csi}$) to suppress noise. To construct accurate fingerprints, we carefully eliminate outliers and suppress noise from the obtained observations. We then systemically evaluate the device uniqueness, time invariance, and mobility independence of micro-CSI-based fingerprints using CSI measurements collected across around 14 months. According to our observations, fingerprint normalization is further adopted to improve the stability of fingerprints.

\item %We subsequently present a robotic use case for our CSI-RFF framework, in which we develop a micro-CSI-based challenge-response authentication algorithm for area access control in mobile robot systems. 
%In industrial environments, certain areas are often restricted due to safety or security concerns. The use of Autonomous Mobile Robots (AMRs) is prevalent in these environments, as they are capable of autonomous navigation and operation. Safety concerns arise due to the potential hazards associated with the use of AMRs. As a result, it is imperative to authenticate mobile robots for area access control to ensure the security of these restricted areas \cite{gavrilova2010state, securingamb}. 
%The CSI-RFF framework applies a simple k-Nearest Neighbors algorithm to authenticate mobile robots based on the WiFi NICs they use, which exactly mimics the use of voice for human identification. For evaluation purposes, 15 COTS WiFi NICs carried by a robot were tested in both static and mobile conditions. The experimental results demonstrate that the CSI-RFF framework can achieve a high attack detection rate (ADR) of over 99\% with a 0\% false alarm rate (FAR) in both static and mobile conditions. These results indicate the effectiveness of the CSI-RFF framework in real-world applications.
We present a practical application of our CSI-RFF framework in a robotic use case, where we devise an authentication algorithm based on micro-CSI for area access control in mobile robot systems. After comparing several state-of-the-art anomaly detection algorithms for open-set authentication, the CSI-RFF framework uses the K-Nearest Neighbors algorithm for authenticating mobile robots, relying on the WiFi NIC they use, an approach analogous to human identification through his/her voice. For evaluation purposes, we conducted tests on 15 COTS WiFi NICs, which were mounted on a robot operating under both static and mobile conditions. Our experimental results showed that CSI-RFF achieved an average attack detection rate (ADR) of close to 99\% while maintaining a 0\% false alarm rate (FAR) across all tested conditions. 

\end{itemize}
Compared to the conference version \cite{kong2023physical}, this work enhances the fingerprint construction process, achieving more robust authentication. Furthermore, it extends the evaluation of CSI-RFF to a mobile robotic use case and broadens the scope to include more advanced WiFi NICs, such as WiFi 4, 5, and 6.

\section{Related Work}
\textbf{Sample-based RF fingerprinting:} RF fingerprinting has been used for both wireless device identification and authentication in recent years. Device identification targets to classify a given device as one of many enrolled devices based on one or more RF fingerprints \cite{oktay,Brik,linning2019,liu,cekic2021wireless,sankhe2019oracle}, while the more challenging device authentication aims to differentiate legitimate and rogue devices under the partial knowledge of the fingerprints of legitimate devices \cite{transient,transient2,hall1,hall2,density,density2,dac,Remley2005,Sheng2008,beam,hua,chen2021securepilot,xie2021generalizable}. Early RF fingerprinting efforts focused on fingerprints extracted from the transient part of transmitted signals \cite{transient,transient2,oktay,hall1,hall2}. Later, other fingerprints were explored, including power spectral density \cite{density,density2}, I/O characteristics of digital-to-analog converter (DAC) and power amplifier (PA) \cite{dac}, electromagnetic signatures \cite{Remley2005}, received signal strength \cite{Sheng2008}, modulation errors \cite{Brik}, beam pattern \cite{beam} and I/Q imbalance \cite{chen2021securepilot}, etc. Recently, there has been wide interest in using deep learning approaches, which learn a richer set of features from raw IQ samples leading to improved performance \cite{cekic2021wireless,sankhe2019oracle,xie2021generalizable}.
All the aforementioned fingerprints need to be extracted from physical-layer samples collected by dedicated and expensive instruments, limiting their direct applications in COTS systems. Furthermore, parts of the research findings were verified by computer simulations \cite{dac,hall1,hall2,cekic2021wireless}, leaving uncertainties on their usability in the real world.

\textbf{CSI-based RF fingerprinting:} We noticed a few works that explored the possibilities of extracting RF fingerprints from CSI measurements of WiFi connections \cite{hua,liu,lin2020}. These works are close to this work, as they also exploited the hardware imperfections incorporated in CSI as fingerprints. Specifically, the authors in \cite{hua} extracted carrier frequency offset (CFO) from CSI measurements and achieved a high rogue access point (AP) ADR of 97.24$\%$ with the FAR being 1.47$\%$ when conducting experiments using 8 commercial APs. However, the CFO fingerprint used in \cite{hua} may not work in mobile scenarios, limiting the authentication accuracy for systems with device mobility. Later, the authors in \cite{liu} extracted nonlinear phase errors from CSI measurements and achieved up to 97$\%$ accuracy for the device classification of 30 WiFi devices. In fact, the multipath effect of wireless channels also contributes to the nonlinear phase. However, authors in \cite{liu} did not separate the nonlinear phase induced by multipath and hardware imperfections. Besides, such a high classification accuracy was achieved by using fingerprints collected in the same small room for both training and testing. Authors in \cite{lin2020} extracted power amplifier (PA)-induced power variance from CSI measurements and combined it with frame interval distribution fingerprint from the network layer to constitute the fingerprint, which can achieve an overall 96.55\% ADR and a 4.31\% FAR when authenticating 12 commercial APs. Compared to \cite{lin2020}, our CSI-RFF, using CSI-carried features only as the fingerprint, can achieve higher than 99\% ADR and 0\% FAR in both static and mobile conditions while using $100$x less CSI measurements. This means that our scheme can substantially decrease the FAR while using fewer CSI measurements and maintaining a slightly higher ADR.

%\textbf{CSI-based position fingerprinting:} It is also worth mentioning another line of research \cite{csi-liu,csi-xiong,csi-jiang,gil2017} that leveraged CSI itself, instead of CSI-carried  RF fingerprints, to realize device authentication. The underlying principle of these works is that CSI values will change if the device location changes more than half a wavelength or the surrounding environment changes. References \cite{csi-liu,csi-jiang} used the correlation of CSI measurements collected in the same position as fingerprints to identify legitimate devices. However, these schemes require legitimate devices to remain static or frequently update CSI measurements of new locations. Mechanisms proposed by \cite{csi-xiong,gil2017} allows the mobility of legitimate devices as they adopt the location information (e.g., angle of arrival) extracted from CSI as fingerprints. These systems require information on legitimate devices’ locations and do not form unique fingerprints for multiple devices. Overall, this line of works fingerprinted wireless channels instead of RF circuitry imperfections of wireless transmitters. Their principles are thus fundamentally different from ours.

\textbf{Artimetrics:} The field of artimetrics involves the identification and authentication of robots, software, and virtual reality agents \cite{yampolskiy2012artimetrics}. %To date, very few works have dealt with the visual or behavioral authentication of robots. The need for the development of robotic biometrics has been identified in \cite{yampolskiy2012artimetrics,gavrilova2010state}.
In \cite{abu2018robot}, a system is proposed for robot authentication using near-field communication (NFC) adapters to ensure secure access control. This system utilizes a unique encryption technique that balances high security with low computational overhead.  On the other hand, research on appearance-based robot detection \cite{kaufmann2005visual,ruiz2010play,bousnina2012learning} often necessitates image inputs and is closely associated with artimetrics. The methodologies proposed in these studies have proven successful in applications such as robotic soccer. Our approach can supplement these encryption- or appearance-based robot identification methods, enhancing their reliability. Particularly, our method remains effective even in scenarios where encryption is compromised or when robots share identical appearances.

\section{Preliminaries and Observations} \label{s3}
In this section, we first explain how the CSI is estimated in commercial WiFi systems and then discuss our initial observations of micro-CSI (i.e., micro-signals that appear on CSI amplitude and phase curves), which can be used as a new RF fingerprint for device authentication.

{Wireless signals experience power attenuation, scattering, reflection, and diffraction during over-the-air propagation. These effects will cause power losses and multipath distortions, with the latter further leading to frequency selectivity over wireless channels. To cope with frequency selectivity, orthogonal frequency division multiplexing (OFDM) technology has been widely adopted in WiFi systems. OFDM employs multiple orthogonal and narrowband carriers, so-called subcarriers, to transmit multiple data symbols at the same time so that it can effectively combat frequency selective fading. 
%Furthermore, OFDM can dramatically simplify channel equalization processes at receiver side through allowing per-subcarrier equalization. 
In wireless communication systems, CSI refers to channel-induced amplitude and phase distortions in the frequency domain. In OFDM WiFi systems, each subcarrier may experience different distortions, and thus CSI estimation needs to be conducted on a per-subcarrier basis. CSI is estimated based on the shared knowledge of reference sequences at the transmitter and receiver sides in WiFi systems \cite{book80211}. %In WiFi systems, the reference sequence refers to the Long Training Field (LTF) in the preamble section of each data packet, which is used for both fine CFO estimation and CSI estimation \cite{book80211}. %The CSI can be estimated by comparing the received reference sequences with the transmitted version, which can be realized by applying various estimation algorithms such as minimum mean-square error (MMSE) or Least-squares (LS) \cite{jan1995}.}

During signal transmission and reception, besides wireless channels, RF circuits may unintentionally alter the transmitted reference sequences and the received sequences because of hardware imperfections, and these alterations are often referred to as RF distortions \cite{book80211}. As the CSI estimation module relies on the standard and unaltered reference sequences to perform the estimation, the estimated CSI values contain not only the distortions caused by wireless channels but also the distortions induced by transceiver RF circuits. Besides, the acquisition of CSI does not need extra resources and/or equipment, since it has been available at the physical layer of wireless communication systems. Motivated by these facts, a few efforts have been made to extract certain aspects of RF distortions from CSI for device identification or authentication. More specifically, RF distortions originating from manufacturing errors can be different for different devices, making it possible to leverage them to realize device authentication. Exemplary RF distortions extracted from CSI measurements include CFO between transceivers \cite{hua}, nonlinear phase errors \cite{liu}, and PA power variances \cite{lin2020}. In the next subsection, we will elaborate on our observations of a new RF distortion embedded in CSI curves and its potential for device fingerprinting.

% \vspace{-1em}
\subsection{Observations of Micro-CSI}
A new CSI-based RF distortion was incidentally observed when we used an Atheros AR9580 NIC equipped with two antennas as the receiver to collect CSI measurements of an Atheros AR9382 NIC equipped with one antenna via the Atheros CSI tool \cite{atheros}. In our experiment, we set the system to work in a bandwidth of 20 MHz, having 56 subcarriers. Figs. \ref{variation}a,b plot the amplitude and phase curves of CSI measurements of the same WiFi card at different times and positions, in which each curve represents one CSI measurement that was collected under different conditions (i.e., three different times in one day and three different positions in the same room). We can see from these figures that some micro-signals (i.e., small variations) appear on the CSI curves of all measured cases. Furthermore, the micro-signals on CSI curves (micro-CSI) look random when we observe them at this single-measurement granularity. This observation made us believe that micro-CSI could be caused by random noise. 
%At this point, we remark that CSI measurements reported by another popular CSI tool, Linux 802.11n \cite{linux}, does not have such apparent micro-signals, confirmed by the CSI curves showed in \cite{liu,picoscenes}, which were generated by that tool. 
Besides, the Atheros CSI tool is only available for AR93/94/95 series of Atheros NICs as the tool was developed based on the Ath9k (ar9003) driver \cite{ath9k}. To figure out whether the micro-CSI is caused by random noise, multipath effects, imperfections of the Atheros CSI tool, or other factors such as RF distortions, we need to conduct more comparison experiments.

\begin{figure}
\centering

 \subfloat[CSI was collected at three times.]{\includegraphics[width=0.5\linewidth]{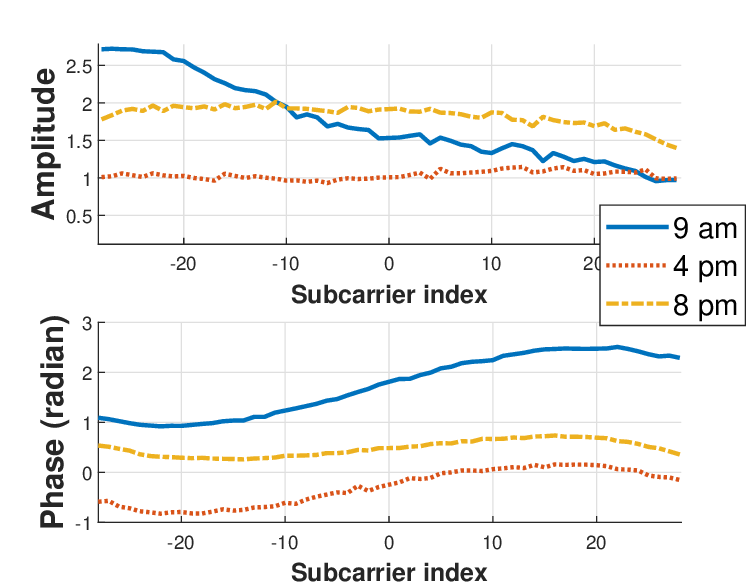}}
\hfil
\subfloat[CSI was collected at three positions in the same room.]{\includegraphics[width=0.5\linewidth]{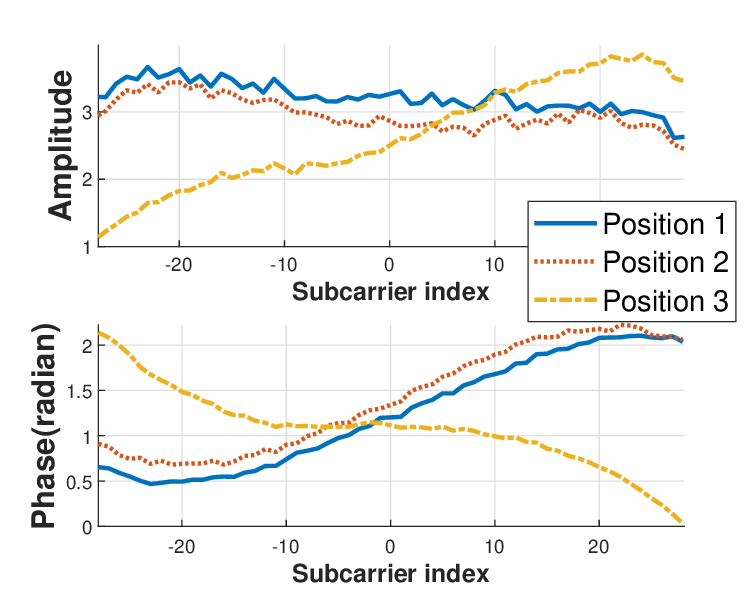}}
\hfill
\vspace{-1em}
\subfloat[Different colored curves represent averaged CSI of different measurement groups, where each group includes 100 consecutive CSI measurements.]{\includegraphics[width=\linewidth]{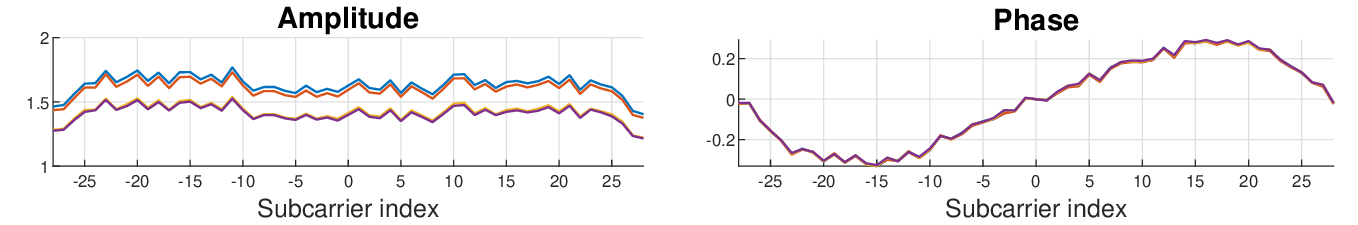}%
\label{stable}}
\caption{Micro-signals shown on CSI curves collected at different times and positions, and a similar pattern of micro-signals appears in each curve after suppressing the effects of random noise. }%(a) CSI was collected at three times for one day. (b) CSI was collected at three positions in the same room. (C) Different colored curves represent averaged CSI of different measurement groups, where each group includes 100 consecutive CSI measurements.}
\label{variation}   
\vspace{-1em}
\end{figure}

First, to verify whether micro-CSI originates from random noise or channel multipath effects, we tried to suppress noise impacts by averaging over multiple replicate CSI measurements of the same WiFi NIC. To do so, we set an AR9382 NIC to send packets and use an AR9580 NIC with the Atheros CSI tool to collect CSI measurements, and the transceiver is connected by an RF cable of 20cm and an attenuator of 30 dB. By using the cable, we can safely assume that the channel is without multipath effects and keeps unchanged so that the collected CSI can be treated as replicate measurements. We started by dividing each 100 consecutive CSI measurements into one group and calculating the averaged CSI amplitude and averaged CSI phase of each group. The averaged CSI amplitude and phase of each measurement group are presented in Fig. \ref{stable}, in which each curve corresponds to one group. We noticed that the micro-CSI on the averaged CSI curves of different groups become relatively stable and exhibit highly similar patterns. This experiment indicates that micro-CSI does not originate from random noise or multipath effects and becomes relatively stable for a transceiver RF chain pair when we suppress noise impacts by averaging the CSI measurements.

We subsequently figured out whether micro-CSI is introduced by the imperfections of the Atheros CSI tool or the RF distortions of WiFi chipsets. In fact, micro signals can also be observed on the CSI curves presented in other works \cite{picoscenes,wolfgang}, which were collected by other tools (e.g., software-defined radio). Besides, we used a set of software-defined radio (SDR) devices to collect CSI of different-brand NICs, see Section \ref{tool} for more details. We observed that the micro-CSI shows up on all tested WiFi 4/5/6 NICs. Due to space limitations, we cannot provide the corresponding CSI curves here. The extracted micro-CSI of several tested NICs is shown in Fig. \ref{unique} later. After observing these CSI curves, we can safely deduce that micro-CSI is not caused by the imperfections of the Atheros CSI tool, and is most likely induced by the RF distortions of WiFi chipsets. %Nevertheless, our results cannot explain why the CSI curves generated by the Linux 802.11n tool do not have obvious micro-CSI. One possible reason is that Intel 5300 chipsets apply smoothing filters before reporting the CSI, as briefly mentioned during the discussions of Fig. 8 in \cite{wolfgang}. But we have no means to confirm this as the details of the Intel 5300 chipset are not public. 

In summary, we have made the following preliminary observations and deductions: (1) Micro-CSI is highly related to RF distortions; (2) Micro-CSI becomes relatively stable on the CSI measurements after suppressing noise by averaging. These properties of the micro-CSI make it can be used as a new viable RF fingerprint for authenticating WiFi devices.% In the following sections, we will systematically evaluate the device-uniqueness, time-invariance, and mobility-independence properties of the micro-CSI-based fingerprint to further confirm its usability as a fingerprint for physical-layer authentication.

\section{Micro-CSI Extraction}\label{s4}

Reliably extracting RF fingerprints from CSI measurements is a non-trivial task as CSI measurements incorporate distortions from both hardware and fluctuating channels. Furthermore, the quality of signal acquisition significantly influences the efficacy of fingerprint extraction \cite{denev}. As such, in the following subsections, we first discuss our considerations for determining the most favorable configurations of WiFi devices during the signal (i.e., CSI measurements) acquisition stage. Subsequently, we explain how to acquire CSI measurements from ACK packets that satisfy our requirements. Finally, we elucidate the signal space-based methodology employed in the extraction of micro-CSI from the CSI measurements.

% \vspace{-1em}
\subsection{Signal Acquisition}
\label{signalAcquisition}
Signals are transmitted from pending authenticated devices and contain unique signal properties (i.e., fingerprints) for device authentication. The acquisition step should retain the unique fingerprint properties that the authentication relies on rather than influencing or degrading the fingerprints, e.g., by introducing measurement errors \cite{denev}. Under this requirement, when it comes to our CSI-based RF fingerprinting system, we need to figure out which configuration is most favorable for collecting CSI measurements for precise fingerprint extraction.

Recent WiFi systems adopt multiple-input multiple-output (MIMO) technologies on top of OFDM to meet the expectation of higher and higher data rate requirements, in which multiple antennas are used to enable the transmission and/or reception of multiple spatial data streams at the same time. In MIMO-OFDM systems, space-time block coding (STBC) encoder at the transmitter and decoder blocks at the receiver are used to support the concurrent transmission of multiple orthogonal spatial streams \cite{stbc}. By applying orthogonal STBC, the system can transmit multiple reference sequences simultaneously and estimate the CSI of all transceiver links at the same time.

\begin{table}%[h]
  \caption{WiFi Devices}
  \label{devices}
  \centering
  \resizebox{0.9\linewidth}{!}{
  \begin{tabular}{ccccc}
    \hline
    NIC No.&WiFi&Brand&Model&Quantity\\
    \hline
    C1-C5&WiFi 4 & Espressif & ESP32-S3 & 5\\
    C6-C9&WiFi 4 &Atheros& AR9271& 4\\
    C10 &WiFi 5& Realtek& RTL8812BU& 1\\
    C11 &WiFi 5& Intel& AC7260& 1\\
    C12 &WiFi 5& Intel& AC7265& 1\\
    C13 &WiFi 5& Intel& AC8260& 1\\
    C14-C15&WiFi 6 & Intel& AX200 & 2\\
  \hline
\end{tabular}
}
\vspace{-1em}
\end{table}

However, in MIMO-OFDM systems, different transmitter chains use separate hardware circuits, thus they may have different fingerprints and the fingerprints of different transmitter chains do not enjoy such orthogonality. 
As such, the MIMO configuration could mix the fingerprints of multiple transmitter chains together, making the fingerprints of different WiFi NICs look more similar. In this regard, we intentionally opt for a single-antenna configuration at the transmitter side during signal acquisition to gather more stable and distinct micro-CSI in various environments. It is noteworthy that many IoT devices utilize a single-antenna configuration and are extensively deployed in smart homes and factories \cite{al2015internet,chen2021securepilot}. On the other hand, our experiments revealed that all multi-antenna NICs tested, as outlined in Table \ref{devices}, employ a fixed single transmitting antenna for sending acknowledgment (ACK) packets. This suggests that the CSI of ACK packets is ideally suited for micro-CSI extraction when working with multi-antenna NICs. Alternatively, we can dynamically adjust the number of ``active" chains of WiFi NICs using appropriate application layer commands or by modifying the NIC driver \cite{picoscenes}. In conclusion, the acquisition of single-antenna CSI poses no significant challenges.
%We remark that such single-antenna configuration is only needed when a device needs to go through the authentication process occasionally. It will not hinder the use of multiple antennas for normal data communications.

 \begin{figure}%[h]  
    \centering
  \includegraphics[width=\linewidth]{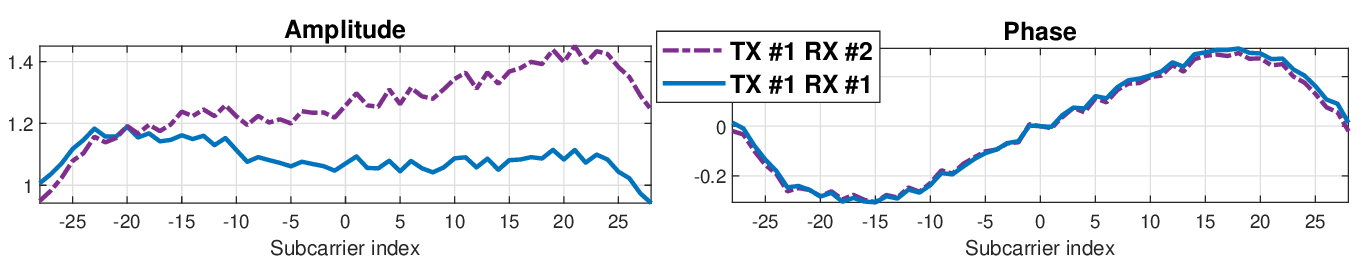}
  \caption{Amplitude and phase curves of the CSI received by different receiver ports of one NIC, where each curve is averaged over $100$ consecutive CSI measurements, which are collected under static conditions with minor CSI changes. Micro signals on each pair of amplitude and phase curves keep almost unchanged.}
  \label{rxchains}
  \vspace{-1em}
\end{figure}

We also conducted an experiment to verify the influence of different receiver chains of a device on micro-CSI. In this experiment, we used two MIMO NICs (AR9382) as the transceiver. Specifically, the transmitter was configured to use a single antenna, and the receiver used two antennas. Fig.~\ref{rxchains} shows that despite the wireless channels between transmitter chain $\#1$ with receiver chains ($\#1,\#2$) being different, receiver chains of the same device cause insignificant differences in micro-CSI. Thus, the SIMO (single-input multiple-output) configuration allows for the collection of CSI of multiple transceiver links carrying the micro-CSI of the single-antenna transmitter at the same time. That means from each CSI measurement, we can extract multiple micro-CSIs, which is beneficial to suppressing noise. Considering this benefit, we used SIMO configuration in all the results presented hereafter.   

% \vspace{-1em}
\subsection{ACK CSI Toolkit}
\label{tool}

To evaluate fingerprints of NICs of different brands and different models, we set up a set of software-defined radio (SDR) devices for CSI collection because all public CSI tools \cite{atheros,linux,nexmon,picoscenes} are not capable of reporting CSI from ACK packets\footnote{A very recent work \cite{esp} validated the extraction of CSI from ACK packets using the commodity ESP32 platform.}. The used SDR consists of a Xilinx Zynq ZC706 development \cite{zc706} and an FMCOMMS5 ADI daughter board with four antennas \cite{coms5}. We use the SDR to collect I/Q samples of packets from different transmitters and process the collected samples for extracting CSI reports\footnote{The CSI Toolkit acts merely as a passive reporter of CSI, with the actual estimation rooted in the physical layer algorithms. The CSI estimation algorithm employed in our framework is the least squares method.} by modifying an 802.11 program provided by MATLAB \cite{analysis}. 

% \vspace{-1em}
\subsection{Signal Space-Based Micro-CSI Extraction}\label{fingerprint_extrac}

Each CSI data comprises information about the wireless channel, micro-CSI induced by hardware distortions, and noise. As we can see from Figs. 1 and 2, wireless channel information and micro-CSI entangle in the frequency domain, making the construction of channel-independent fingerprint non-trivial. In the following, we first define the signal models of the channel information and micro-CSI and then present a signal space-based micro-CSI extraction method for line-of-sight (LoS) conditions.

The estimation of CSI in 802.11 protocols is accomplished based on the Long Training Symbol (LTS) included in the preamble part of each packet. To account for interference from analog circuits and neighboring bands, commodity WiFi systems only use a subset of subcarriers for communication, denoted by $\mathcal{K}$. Denote by ${\bold{t}} = [{t}_0, {t}_1,\dots,{t}_{N-1}]^T$ and $\tilde{\bold{t}} = [\tilde{t}_{-|\mathcal{K}|/2},\dots,\tilde{t}_{|\mathcal{K}|/2}]^T$ the LTS in the time domain and frequency domain, respectively. Here, $N$ is the length of the Discrete Fourier Transform (DFT). The transmitter's introduced hardware distortions (i.e., micro-CSI) in the time and frequency domains are respectively represented by $\bold{d} = [d_0,d_1,\dots,d_{N-1}]^T$ and $\tilde{\bold{d}} = [\tilde{d}_{-|\mathcal{K}|/2},\dots,\tilde{d}_{|\mathcal{K}|/2}]^T$. We further have $\tilde{\bold{t}} = \bold{F}_{\mathcal{K},\mathcal{N}}\bold{t}$ and $\tilde{\bold{d}} = \bold{F}_{\mathcal{K},\mathcal{N}}\bold{d}$, where $\bold{F}$ is the full unitary DFT matrix, and $\bold{F}_{\mathcal{K},\mathcal{N}}$ is the sub-matrix of $\bold{F}$, comprising all rows with indexes in $\mathcal{K}$ and all columns with indexes in $\mathcal{N}$, where $\mathcal{N} = \{0,\dots,N-1\}$. For example, for the transmission of the LTS of ACK packets in all 802.11 protocols, a 64-point DFT is adopted, with only 52 subcarriers being used  (i.e., $|\mathcal{K}| = 52$). The index sets used in the time domain and frequency domain are defined as $\mathcal{N}=\{0,\dots, 63\}$ and $\mathcal{K}=\{-26,-25,\dots,-1,1,\dots,26\}$, respectively. 

Mathematically, we can regard the hardware distortions as some deviations made to the standard LTS samples. Under this model, the time-domain signal emitted to the air can then be written as $\bold{s=t+d}$. Thanks to the cyclic prefix (CP) adopted in OFDM systems, the received signal can be written as $\bold{y=h \ast s + z}$, where $\ast$ denotes circular convolution, $\bold{h}$ is the discrete-time equivalent channel, and $\bold{z}$ is complex white Gaussian noise. After synchronizing the received samples, the CSI in the frequency domain can be estimated by least squares (LS) estimation \cite{book80211}. Specifically,  
\begin{equation}\label{est_CSI}\tilde{\bold{c}}=\tilde{\bold{y}}\circ\tilde{\bold{t}}=\tilde{\bold{h}}\circ(\bold{1}+\tilde{\bold{d}}\circ\tilde{\bold{t}})+\tilde{\bold{z}} = \tilde{\bold{h}}+\tilde{\bold{h}}\circ \Tilde{\bold{f}}+\tilde{\bold{z}} ,
\end{equation}
where  $\Tilde{\bold{f}}=\Tilde{\bold{d}}\circ\Tilde{\bold{t}}$ can be treated as the micro-CSI, $\circ$ is element-wise multiplication and $\bold{1}$ represents the vector with all 1's, and $\tilde{\bold{z}}$ is the frequency-domain noise vector with each element $\tilde{z}_k$ follows a complex normal distribution.%, $\tilde{z}_k\sim\mathcal{CN}(0,\sigma^2)$. 
We can see from (\ref{est_CSI}) that the channel information $\tilde{\bold{h}}$ and the hardware imperfection $\tilde{\bold{d}}$ entangle together. To circumvent the problem, our basic idea is to analyze the CSI measurements from the signal space perspective. Specifically, thanks to the inherent channel sparsity, the number of taps of wireless channels in the time domain (i.e., $\bold{h}$) is often much smaller than the DFT length $N$. As such, the micro-CSI that resides in other unoccupied dimensions can be separated from the channel information. As the first attempt to solve the problem, we start with the following scenarios with strong LoS.

\begin{figure}%[h]  
    \centering
\includegraphics[width=\linewidth]{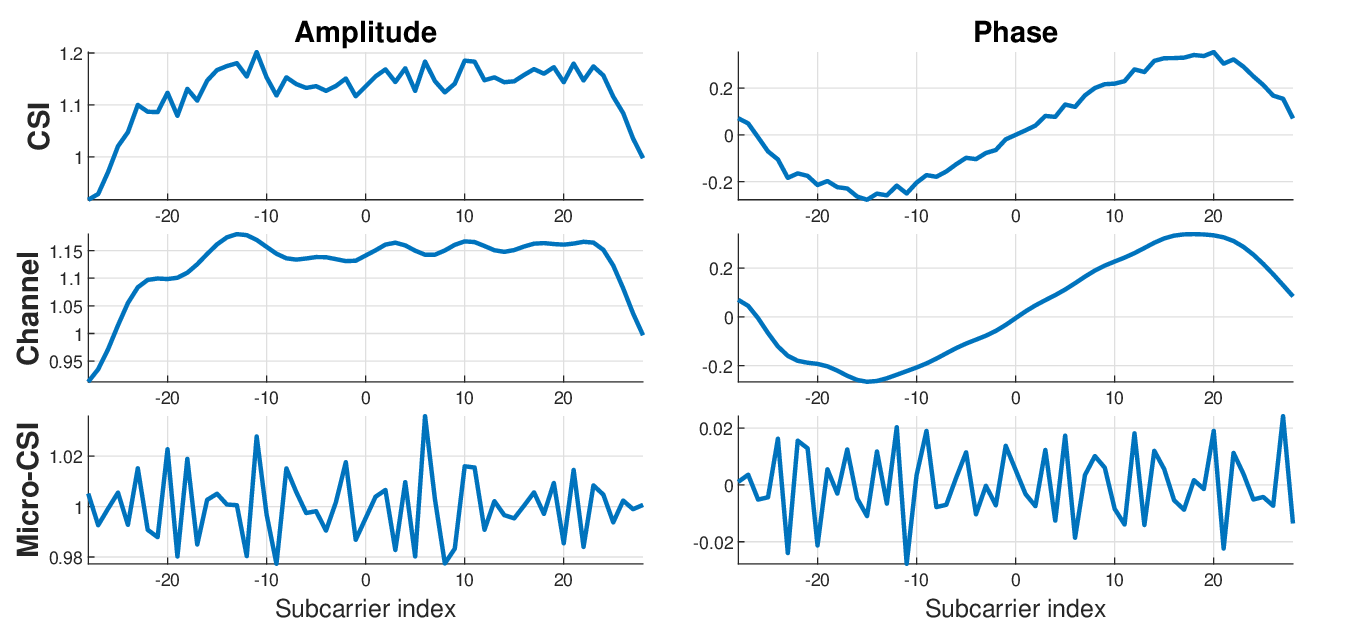}
  \caption{Extracted Micro-CSI from practical CSI measurement.}
  \label{fingerprint}
   \vspace{-1em}
\end{figure}

Under strong LoS, we can safely assume that the first arrived physical path has the strongest power and dominates the channel response. However, a single physical path does not mean that the time-domain CSI $\bold h$ only has one tap. This is because a single path can lead to multiple channel taps when the propagation delay is not an integer multiple of the sampling rate $T_s$, which is known as the shaping/channel leakage \cite{bala2013shaping}. The number of leakage taps depends on the pulse-shaping filter, accounting for leakage on both earlier and later taps. In 802.11 protocols, frame synchronization is achieved using correlation-based algorithms that align with the strongest signal tap, which is also the central tap in strong LoS scenarios. Besides, taps before the strongest tap will be circularly shifted by $N$. Therefore, the set of taps that carries channel information is $\mathcal{L} =\{-N_p,..., N_p\} \bmod N$, where $N_p$ is the number of leaked taps on each side. An example of a single-path leaked channel is given in \cite[Fig. 1]{xk_spawc}. Due to the rapid decay of the impulse response of the pulse shaping filter, we set  $N_p=8$ in all our experiments. The underlying assumption is that the pulse shape will decay to zero after 8 sampling time intervals. Later in Section \ref{s5}, we will evaluate how the value of $N_p$ used in our algorithm affects the accuracy of the device authentication performance. 

%Let $\alpha$ denote the attenuation of the strongest path that dominates the channel, $\tilde{\bold{h}}_p=[\tilde{h}_{p,-|\mathcal{K}|/2},\dots,\tilde{h}_{p,|\mathcal{K}|/2}]^T$ denote the channel leakage effect in the frequency domain. The effect of channel leakage depends on the pulse shaping filter and fractional path delay. 
%The discrete-time equivalents channel $\tilde{\bold{h}}=\alpha \tilde{\bold{h}}_p$ captures the joint effects of channel and filtering.  
%Under strong LoS conditions without STO, we can safely assume that $M=1$ and $h_{l\bmod N}= \alpha p(lT_s)$. We further have $\mathcal{L} =1$, $\bold{{h}} = {h}_1=\alpha$, and $\tilde{\bold{h}} = \bold{F}_{\mathcal{K},\mathcal{L}}\bold{h} = \alpha \bold{1}$, which are constant vectors dependent only on the attenuation value of the first path. 
% Therefore, the estimated CSI under strong LoS conditions can be approximated as
% \begin{align}
% \bold{\tilde{c}} &\approx \alpha \tilde{\bold{h}}_p+\alpha \tilde{\bold{h}}_p \circ\bold{\tilde{f}},
% \end{align}                    
% where $\tilde{\bold{f}}=\tilde{\bold{d}}\circ\tilde{\bold{t}}$ can be treated as the fingerprint, we ignore noise for simplicity. %In this context, the only term that varies with channel conditions is $\alpha$.
To construct a channel-independent fingerprint, we need to eliminate the channel $\Tilde{\bold{h}}$ from the estimated CSI $\bold{\tilde{c}}$. 
Under strong LoS, we can assume that each element of the term $\Tilde{\bold{h}}$ is much larger than that of the term $\Tilde{\bold{h}}\circ\bold{\tilde{f}}$, considering the small scale of $\bold{\tilde{f}}$. This indicates that the values on these time-domain taps occupied by the channel can be considered to be solely contributed by the channel information. In this context, leveraging the fact that the number of channel taps is limited, we can apply the LS method \cite{ls} to estimate the frequency-domain channel information by
\begin{align}
    \hat{\tilde{\bold{h}}}&=\bold{F}_{\mathcal{K},\mathcal{L}}(\bold{F}_{\mathcal{K},\mathcal{L}}^H\bold{F}_{\mathcal{K},\mathcal{L}})^{-1}\bold{F}_{\mathcal{K},\mathcal{L}}^H\tilde{\bold{c}} \approx \Tilde{\bold{h}}
    %&=\alpha \tilde{\bold{h}}_p+\bold{A}(\alpha \tilde{\bold{h}}_p\circ\tilde{\bold{f}}),
%&=\alpha\bold{1}+\alpha\bold{A}\tilde{\bold{f}}+\bold{A}\tilde{\bold{z}},
\end{align}
%where $\bold{A}=\bold{F}_{\mathcal{K},\mathcal{L}}(\bold{F}_{\mathcal{K},\mathcal{L}}^H\bold{F}_{\mathcal{K},\mathcal{L}})^{-1}\bold{F}_{\mathcal{K},\mathcal{L}}^H$ and 
where $(\cdot)^{H}$ and $(\cdot)^{-1}$ represent the Hermitian transpose and the matrix inverse, respectively. The first $\bold{F}_{\mathcal{K},\mathcal{L}}^H\tilde{\bold{c}}$ transform CSI measurement to time domain and keep only the channel taps within $\mathcal{L}$. The second step $(\bold{F}_{\mathcal{K},\mathcal{L}}^H\bold{F}_{\mathcal{K},\mathcal{L}})^{-1}$ is a correct term because $\bold{F}_{\mathcal{K},\mathcal{L}}$ is a partial DFT. The third step $\bold{F}_{\mathcal{K},\mathcal{L}}$ transforms back to the frequency domain. 
Consequently, the micro-CSI can be estimated by element-wise dividing $\tilde{\bold{c}}$ by $\hat{\tilde{\bold{h}}}$. That is,
\begin{equation}
\label{ls}
    \hat{\tilde{\bold{f}}}= \tilde{\bold{c}} ./
\hat{\tilde{\bold{h}}}
%=\frac{ \bold{1}+ \bold{\tilde{f}}}{\bold{1}+\bold{A}(\tilde{\bold{h}}_p \circ \tilde{\bold{f}})./\tilde{\bold{h}}_p} 
\approx \bold{1}+ \bold{\tilde{f}}.%\frac{ \bold{1}+ \bold{\tilde{f}}}{\bold{1}+\bold{A}\tilde{\bold{f}}}.
\end{equation}
%\red{The estimated fingerprint is solely determined by the fingerprint $\bold{\tilde{f}}$ and the pulse shaping filter $\tilde{\bold{h}}_p$, while the matrix $\bold{A}$ remains unchanged. These two aspects are determined by the transmitter. Consequently, despite the presence of biases in the estimated fingerprint, it is device-specific and channel-independent.} 
The estimated micro-CSI is solely determined by the hardware distortions, which is device-specific and channel-independent. 
Fig.~\ref{fingerprint} shows the micro-CSI extracted from one practical CSI measurement. We can see that our algorithm can effectively extract the micro-CSI embedded in CSI.

\section{Micro-CSI-based RF Fingerprinting}
Constructing micro-CSI-based RF fingerprints is a non-trivial task. 
%red{On one side, the signal processing process of receiver chains can introduce additional processing errors to the signals from the transmitter, complicating the recovery of the transmitter's fingerprints}. On the other side, 
This is because RF distortions reflected in the CSI may be weak and susceptible to contamination by noise. In the following subsections, we first elaborate on how to use micro-CSI to construct a new RF fingerprint, considering the impact of processing errors from receiver chains. Subsequently, we evaluate the robustness of micro-CSI-based fingerprints for device authentication in terms of device uniqueness, time invariance, and noise resistance. We finally designed a KNN algorithm-based fingerprint matcher for device authentication.

% \vspace{-1em}
\subsection{Fingerprint Construction}
We now elaborate on how to construct a fingerprint based on a group of extracted micro-CSIs. Recall that $|\mathcal{K}|$ denotes the total number of subcarriers. Let $\hat{\tilde{\bold{f}}}_n \in {\mathbb{C}}^{|\mathcal{K}|\times1}$ denote the $n$-th estimated micro-CSI, where $n \in \mathcal{N}_{m}=\{1,...,|\mathcal{N}_{m}|\}$, and $|\mathcal{N}_{m}|$ denotes the total number of micro-CSIs. Denote by $\hat{\tilde{f}}_{n,k}$ the micro-CSI presented on the $k$-th subcarrier in the $n$-th micro-CSI. 

\subsubsection{Noise Suppresion}
As discussed in Section \ref{s3}, the micro-CSI, due to its small scale, can easily be contaminated by noise. To suppress noise effects, we average among a group of micro-CSIs extracted from CSI measurements. Hereafter, $N_{csi}$ represents the number of CSI measurements used to construct one fingerprint. Later in Section \ref{s5}, we will evaluate how the value of $N_{csi}$ affects the accuracy of the device authentication performance. In $1 \times N_{rx}$ SIMO configuration, each CSI measurement carries $N_{rx}$ micro-CSIs, where $N_{rx}$ denotes the number of receiver chains. As discussed in Section \ref{signalAcquisition}, the RF chains of the receiver cause insignificant differences in micro-CSI, the $N_{rx}$ micro-CSIs can thus be used as if they were collected by the same antenna. Therefore, $|\mathcal{N}_{m}|=N_{rx}\times N_{csi}$.

\subsubsection{Outlier Elimination}
Based on our experiments, we have identified three types of CSI measurements. For example, in one instance, we collected CSI measurements of one Atheros NIC (AR9271) from ACK packets under a static condition, see Section \ref{tool} for more details. In Fig. \ref{outlier}, the left figure depicts the amplitude of three different types of CSI measurements, while the figures in the right column correspond to the respective micro-CSIs. The corresponding phases demonstrate the same phenomenons and their curves are omitted for brevity. Most CSI measurements are similar to the normal case in Fig. \ref{outlier}, where the intensity of micro-CSI is small. However, a few CSI measurements exhibit abnormal behaviors. In abnormal case 1, shown in Fig. \ref{outlier}, the intensity of micro-CSI on a few subcarriers is unusually large. In abnormal case 2, the intensity of micro-CSI is unusually large across all subcarriers. Upon observing the corresponding samples in the time domain, we notice that the cause of abnormal case 1 is the burst high noise in the samples, and the cause of abnormal case 2 is the clipping of samples caused by a large and inappropriate receiver gain. These two abnormal cases were caused by the misbehavior of the CSI collector used or potential interference in real-world environments, but the samples could still be decoded successfully due to the robustness of binary phase shift keying (BPSK) used in ACK. Therefore, we need to eliminate these abnormal cases as they carry inaccurate micro-CSI.

\begin{figure}%[h]  
    \centering
\includegraphics[width=\linewidth]{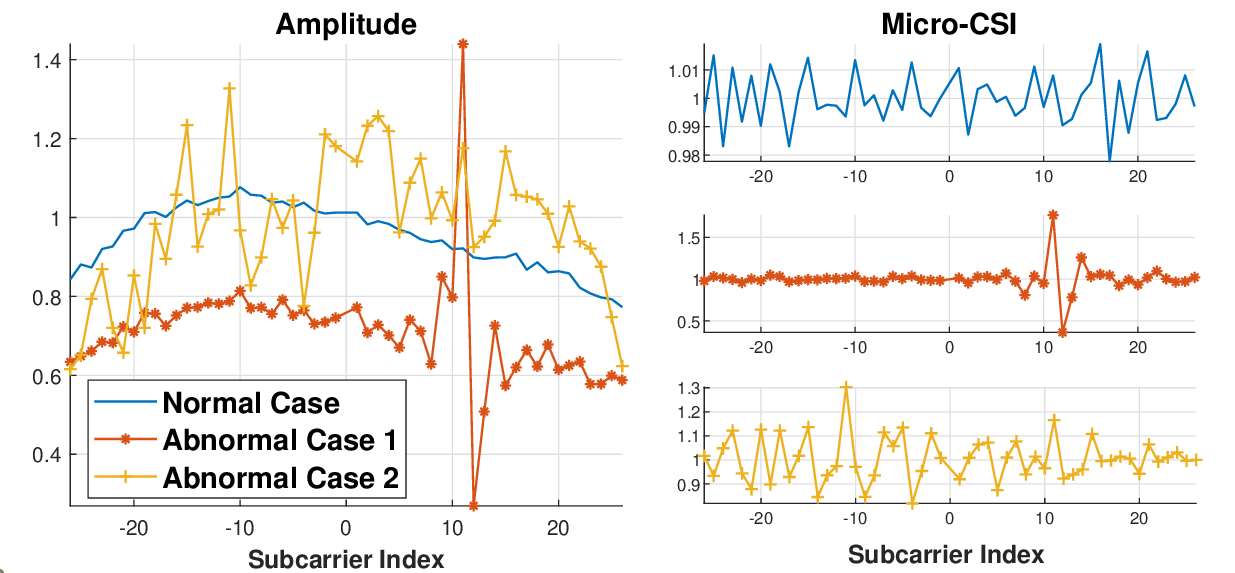}
  \caption{Three types of CSI measurements.}
  \label{outlier} 
   \vspace{-1em}
\end{figure}

We propose two strategies for addressing the two abnormal cases. Firstly, for abnormal case 2, we eliminate it by identifying the large variance of gradients across subcarriers. The gradients across subcarriers can be formulated as:
\begin{equation}
    \nabla \hat{\tilde{\bold{f}}}_n = \{ \hat{\tilde{f}}_{n,1-|\mathcal{K}|/2}-\hat{\tilde{f}}_{n,-|\mathcal{K}|/2},...,\hat{\tilde{f}}_{n,|\mathcal{K}|/2}-\hat{\tilde{f}}_{n,|\mathcal{K}|/2-1} \}.
\end{equation}
Compared with the normal case, gradients in abnormal case 2 change dramatically. Consequently, the variance of gradients in abnormal case 2 is much larger than in the normal case. By setting a threshold for the variance of gradients, we can effectively identify and eliminate abnormal case 2. According to our observations that variances of all micro-CSI in the normal case are below $2\times10^{-3}$, we then set the threshold $\omega_1=2\times10^{-3}$. 

After eliminating the micro-CSI of abnormal case 2, we obtain the remaining $|\mathcal{N}_{m}'|$ micro-CSIs. For abnormal case 1, the rest of the subcarriers with the small intensity of micro signals can still be used to suppress noise. Therefore, we only need to eliminate outliers of the remaining $|\mathcal{N}_{m}'|$ micro-CSIs for each subcarrier. To do so, we adopt $Z$-score \cite{zscore} as the metric to quantify the unusualness of each observation. Specifically, for each subcarrier $k \in \mathcal{K}$, we only keep micro-CSIs of each subcarrier that satisfy
\begin{equation}
\label{zscores}
    \mathcal{Z}(\hat{\tilde{f}}_{n,k})=\left|\frac{\hat{\tilde{f}}_{n,k}-u_{k}}{\sigma_{k}}\right| < \omega_2, n\in\mathcal{N}_{m}'
\end{equation}
where $u_{k}$ and $\sigma_{k}$ are the sample mean and variance of the remaining of $|\mathcal{N}_{m}'|$ micro-CSI values of subcarrier $k$, respectively. 
We set the threshold $\omega_2$ as 1, meaning that we only keep these micro-CSI values that are a standard deviation above/below the sample mean. The determination of the value for $\omega_2$ stems from the consideration that when the sample size is relatively small (capped at 80 in our evaluation), the accuracy of the population mean and standard deviation estimates may be compromised. Lowering the threshold for the z-score allows for a more sensitive detection of outliers in such circumstances. Additionally, the validity of this choice is corroborated by our empirical results. Then, the kept micro-CSI values of subcarrier $k$ that satisfy (\ref{zscores}) are averaged to suppress noise and the denoised fingerprint for subcarrier $k$ is denoted by $\tilde{f}_k^{'}$. Collectively, the denoised fingerprint for all subcarriers is represented as $\tilde{\bold{f}}^{'}=\{\tilde{f}_k^{'}\}_{k\in \mathcal{K}}$. 

We further adopt $Z$-score to normalize the denoised fingerprints to enhance the stability of estimated fingerprints, according to our observations in Section \ref{robustness}. The improved stability of normalized fingerprints is illustrated in Fig.~\ref{stability}.  
The normalized fingerprint that used for authentication is denoted by $\tilde{\bold{f}}^{a} = \mathcal{Z}(\tilde{\bold{f}}^{'})$. We present Algorithm \ref{fc} to provide a clear outline of the process of outliers elimination, noise suppression, and fingerprint normalization.

\begin{algorithm}
 \caption{Fingerprint construction}
 \label{fc}

\LinesNumbered
\KwIn{A group of micro-CSIs of the same device $\hat{\tilde{\bold{f}}}_n$, where $n \in \mathcal{N}_{m}$; Threshold $\omega_1, \omega_2$ }%\\
\KwOut{The denoised and normalized fingerprint $\tilde{\bold{f}}^{a}$}%\\
Initialization: $\mathcal{N}_{m}'=\emptyset$\\

\For{$n \in \mathcal{N}_{m}$ }{ 
    \If{ $\Var(\nabla \hat{\tilde{\bold{f}}}_n) < \omega_1$}{$\mathcal{N}_{m}'\leftarrow n$ \tcp*[r]{abnormal case 2}} 
    }
\For{$k \in \mathcal{K}$}{ 
$\mathcal{F}=\emptyset$\\
    \For{$n \in \mathcal{N}_{m}'$}{
\If{$\left|\frac{\hat{\tilde{f}}_{n,k}-u_{k}}{\sigma_{k}}\right| < \omega_2$}{$\mathcal{F}\leftarrow \hat{\tilde{f}}_{n,k}$ \tcp*[r]{abnormal case 1}}
}
$\tilde{f}_k^{'}=\overline{\mathcal{F}}$ \tcp*{suppress noise}
}
$\tilde{\bold{f}}^{'} = \{\tilde{f}_k^{'}\}_{k\in \mathcal{K}}, u = \overline{\tilde{\bold{f}}^{'}}, \sigma = \Var(\tilde{\bold{f}}^{'})$ \\ %\tcp*[r]{denoised fingerprint}
$\tilde{\bold{f}}^{a} = \{ \frac{\tilde{f}_k^{'}-u}{\sigma} \}_{k\in \mathcal{K}} $ \tcp*[r]{normalized fingerprint} %\mathcal{Z}(\tilde{\bold{f}}^{'})
\Return $\tilde{\bold{f}}^{a} $
\end{algorithm}

\vspace{-1em}
\subsection{Evaluations of Fingerprint Robustness}\label{robustness}

We now present detailed experiments and results to examine the robustness of fingerprints in terms of device uniqueness, time invariance, and noise resistance. We remark that all CSI measurements used in subsequent analyses are obtained from ACK packets by toolkit detailed in Section \ref{tool}. 
\begin{figure*}%[ht]  
    \centering
    \subfloat[Atheros AR9371 $\#1$]{\includegraphics[width=0.23\textwidth]{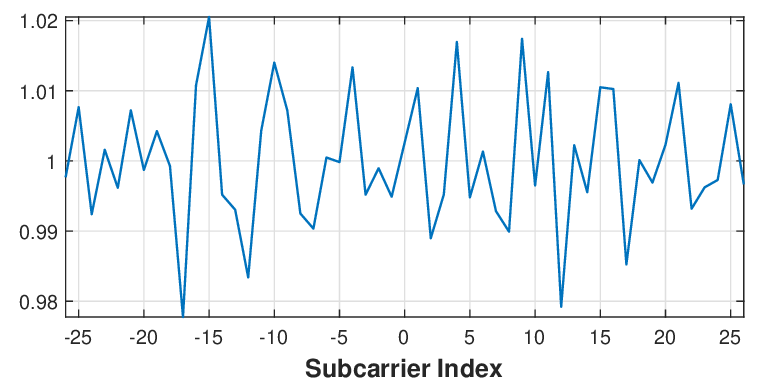} \label{ua}}
    \subfloat[Espressif ESP32-S3 $\#1$]{\includegraphics[width=0.23\textwidth]{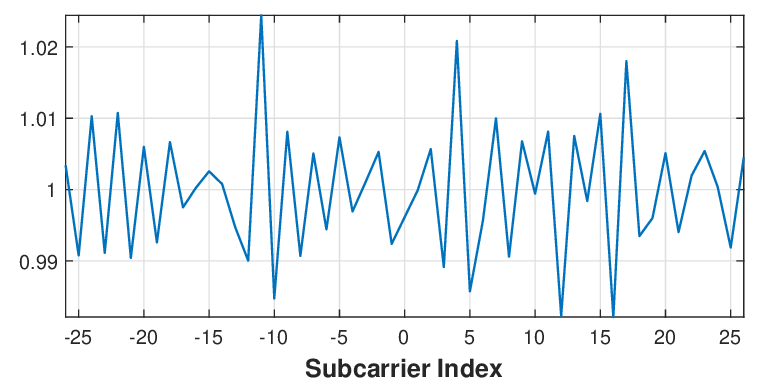} \label{ub}}
    \subfloat[Intel AX200 $\#1$]{\includegraphics[width=0.23\textwidth]{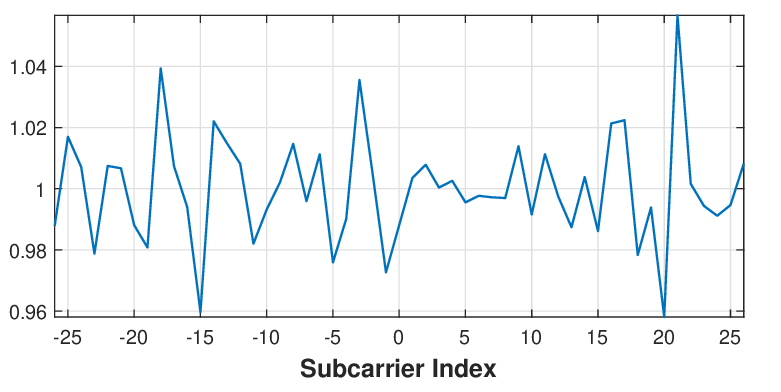} \label{uc}}
    \subfloat[Intel AC7260]{\includegraphics[width=0.23\textwidth]{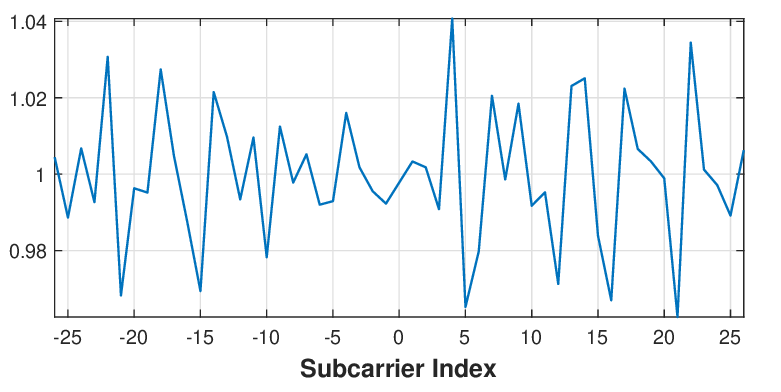} \label{ud}}

    \subfloat[Atheros AR9371 $\#2$]{\includegraphics[width=0.23\textwidth]{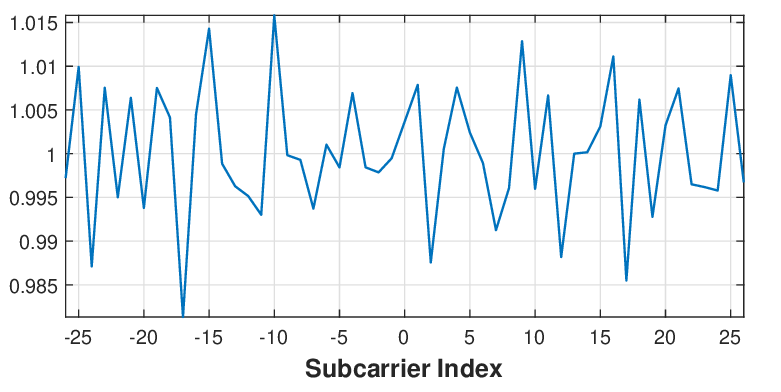} \label{ue}}
    \subfloat[Espressif ESP32-S3 $\#2$]{\includegraphics[width=0.23\textwidth]{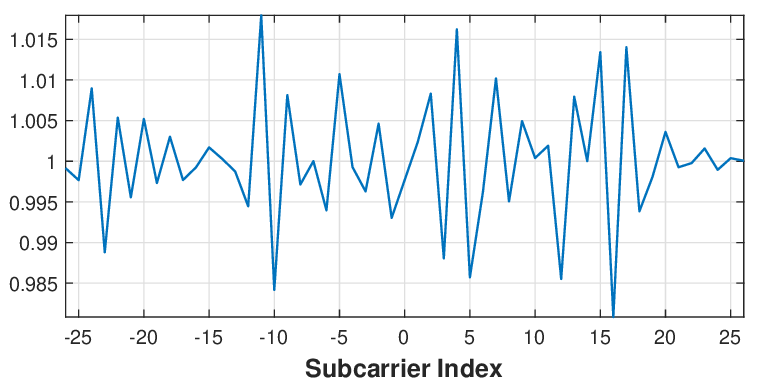} \label{uf}}
    \subfloat[Intel AX200 $\#2$]{\includegraphics[width=0.23\textwidth]{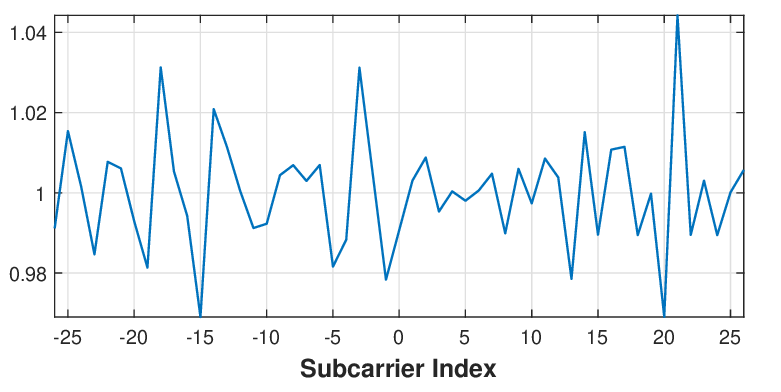} \label{ug}}
    \subfloat[Intel AC7265]{\includegraphics[width=0.23\textwidth]{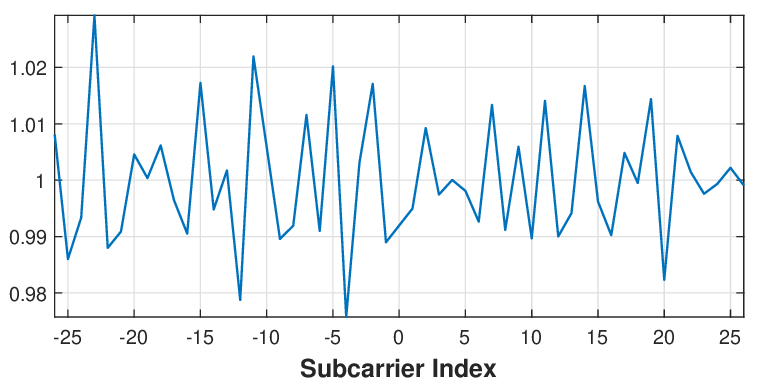} \label{uh}}
  \caption{Uniqueness of amplitude of fingerprints of 8 NICs of various brands, where $N_{csi}=20, N_{rx}=4$.}
  \label{unique}
  \vspace{-1em}
\end{figure*}

\textbf{Device Uniqueness}. An ideal fingerprint needs to be unique and distinguishable from device to device. %To verify the uniqueness of NICs of different brands, different models of the same brand, and the same model, we set up a set of software-defined radio (SDR) devices to collect CSI measurements for analysis because the Atheros CSI tool is only available for Atheros NICs and other CSI tools are not capable to report CSI from ACK packets. 
%The used SDR consists of a Xilinx Zynq ZC706 development \cite{zc706} and an FMCOMMS5 ADI daughter board with four antennas \cite{coms5}. We use the SDR to collect I/Q samples of packets from different transmitters and processed the collected samples for extracting CSI reports by leveraging an 802.11 program provided by MATLAB \cite{analysis}. 
In this experiment, we collected the CSI from the signals transmitted by WiFi 4 (Atheros AR9371, Espressif ESP32-S3), WiFi 5 (Realtek RTL8812BU), and WiFi 6 (Intel AX200) NICs at the same location. These 8 NICs serve as transmitters and are supported by the same host PC, except ESP32-S3 as its WiFi module is embedded in itself. This assured that the fingerprint difference, if exists, is solely caused by the NICs. For each NIC, we collected 20 CSI measurements to construct its fingerprints. {Fig. \ref{unique} plots the amplitude fingerprints of the 8 NICs tested in our experiments. We can see that the fingerprints of different NICs are indeed distinguishable. The phase fingerprints of these devices demonstrated similar device uniqueness, and their curves are omitted for brevity.}

From Figs. \ref{unique}a-d, we can observe that the fingerprints of NICs from different brands are distinctive, and the similarities between them are small. Fingerprints of NICs from different models of the same brand are still distinctive enough, as shown in Figs.~\ref{ud}, and \ref{uh}. For NICs of the same model, we can observe that their fingerprints are still distinguishable while their similarities become higher than the previous two cases when comparing Figs. ~\ref{ua} and \ref{ue}, Figs.~\ref{ub} and \ref{uf}, and Figs.~\ref{uc} and \ref{ug}. In short, the closer the model types of NICs, the more similar their fingerprints are. The reason behind this is that NICs from the same manufacturer are highly probable to experience similar manufacturing imperfections, rendering similar RF distortions. Fortunately, even for NICs of the same model, their fingerprints are still not exactly the same, indicating the potential of micro-CSI-based fingerprints for performing the highly challenging authentication of NICs of the same model. 

% \vspace{-1em}
\textbf{Time Invariance}. An ideal fingerprint needs to be time-invariant, and it should keep stable even when the operating conditions (e.g., temperature) change. To conduct the stability test, we compare two sets of CSI measurements of 2 Realtek NICs (RTL8812BU) collected with a 14-month gap, one set in January 2022 and the other one in May 2023. In each collection, we collected around 200 ACK packets for each NIC in about 0.1 seconds to the CSI collector (SDR), and we averaged 50 micro-CSIs extracted from consecutive CSI measurements to construct one fingerprint of each NIC. Fig. \ref{stability} shows the statistical analysis of the fingerprints of one NIC. The other NIC demonstrated similar stability, and their curves are omitted for brevity. As shown in Fig.~\ref{stability}a, the boxes for most of the indexes are small, which means variances of fingerprints of these indexes are small. We also can observe that fingerprints in the amplitude domain show more variance compared to the fingerprint stability in the phase domain. We then estimate the $Z$-score of each fingerprint, and the statistical analysis is shown in Fig.~\ref{stability}b. The stability of fingerprints improves, which means the larger variances in fingerprint amplitude in Fig.~\ref{stability}a are because of the slight variance in the intensity of the fingerprint. Therefore, this motivates us to normalize the amplitude of the fingerprint by using $Z$-score in Algorithm \ref{fc} to ensure the fingerprint across subcarriers has zero mean and the standard deviation is 1. 
%this indicates that our training data in the evaluation should cover more variants of fingerprints to achieve better performance. %This is the reason why training with mobile data shows better authentication performance than only using static data, as discussed in Section 5.

\begin{figure*}
\centering
 \subfloat[Stability of the fingerprints with 14 months’ gap.]{\includegraphics[width=0.5\linewidth]{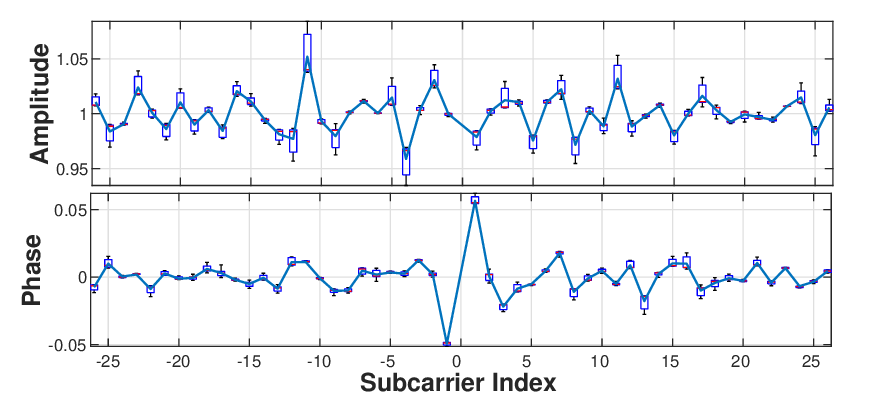}}
 \hfill
\subfloat[Stability of the $Z$-scores of the fingerprints.]{\includegraphics[width=0.5\linewidth]{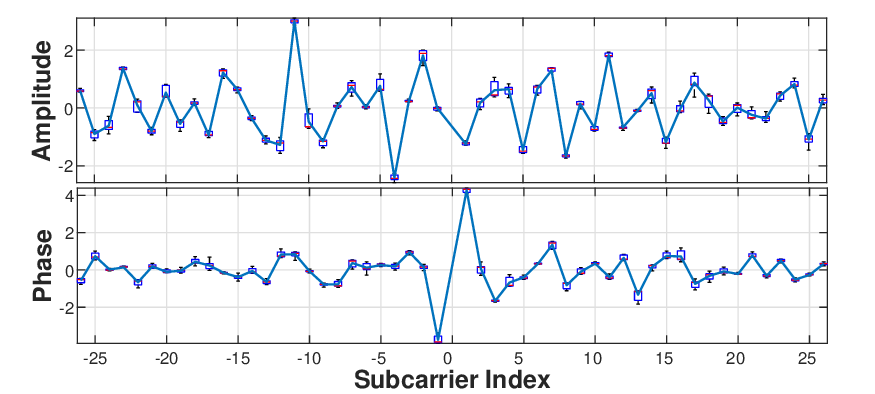}}
\caption{Invariance of the fingerprints with respect to time, where $N_{csi}=50,N_{rx}=1$. }
\label{stability}   
\vspace{-1em}
\end{figure*}

% \begin{figure*}%[h]
%   \centering
%   \begin{minipage}[]{0.6\textwidth}
%   \subfloat[Best Case.]{\includegraphics[width=0.5\linewidth]{figures/distance_ac.eps}}
%   \hfil
%   % \vspace{-0.5em}
%   \subfloat[Worst Case.]{\includegraphics[width=0.5\linewidth]{figures/distance_esp.eps}}
%   \caption{Influence of noise on the needed number of CSI measurements $N_{csi}$, where $N_{rx}=4$.} 
%   \label{distance}
%   \end{minipage}
% \hfill
%   \begin{minipage}[]{0.36\textwidth}
%       \includegraphics[width=\linewidth]{figures/mobility_compare.eps}
%   \caption{Mobility influence on fingerprints, where $N_{csi}=100, N_{rx}=1$.}
%   \label{mobility}
%   \end{minipage}
%   \vspace{-1em}
% \end{figure*}

\begin{figure}
\centering
\subfloat[Best Case.]{\includegraphics[width=\linewidth]{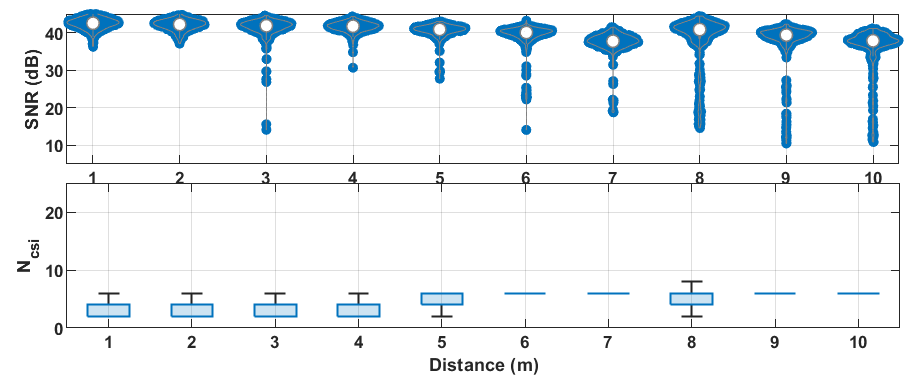}}
\hfil
\vspace{-0.1em}
\subfloat[Worst Case.]{\includegraphics[width=\linewidth]{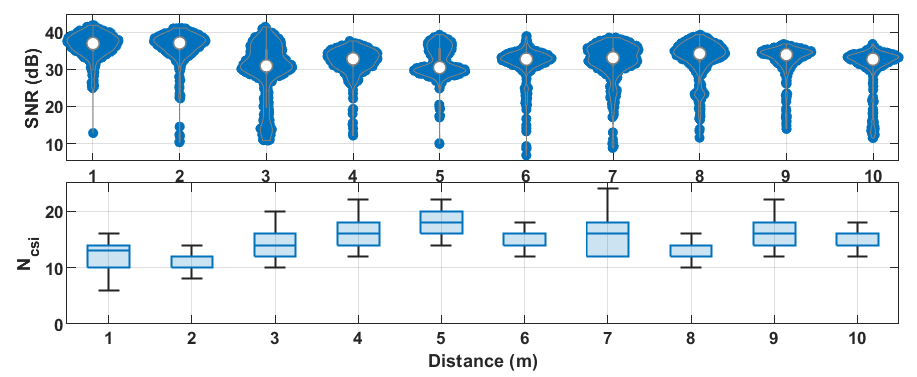}}
\caption{Influence of noise on the needed number of CSI measurements $N_{csi}$, where $N_{rx}=4$.} 
\label{distance}
\end{figure}

% \vspace{-1em}

\textbf{Noise Resistance.} 
To address the impact of noise on our fingerprinting approach, we average among a group of micro-CSIs extracted from CSI measurements. We have conducted experiments to determine the suitable number of CSI measurements ($N_{csi}$) necessary to achieve reliable fingerprinting over LoS distances from 1 to 10 meters between transceivers. We have implemented an iterative convergence strategy to specify the number of CSI measurements needed. Our approach ensures that the number of CSI measurements used is sufficient to yield a stable and accurate fingerprint, which we define as the point at which consecutive estimates of the fingerprint do not differ by more than a predefined threshold. The difference between consecutive estimates of the fingerprint is defined as relative error as $e=\frac{||\{\tilde{\bold{f}}^a\}^{i+1}-\{\tilde{\bold{f}}^a\}^{i}||_2}{||\{\tilde{\bold{f}}^a\}^{i}||_2}$, where $i$ is the iteration index. Based on empirical analysis, we have set this threshold at 0.1. The initial value for $N_{csi}$ is set to 1 and is increased by 2 in each iteration until the desired stability is achieved.

Illustrated in Fig.~\ref{distance} are the signal-to-noise ratio (SNR) values associated with the collected CSI measurements at varying distances, as well as the required number of CSI measurements ($N_{csi}$) for two NICs, which exemplify the best (RTL8812BU) and worst-case (ESP32) scenarios. A higher transmitting power and a more pronounced fingerprint are considered indicative of a better case for noise suppression. Within SNR estimation, the signal power is estimated by the mean square of the absolute values of training symbols, and noise power is equal to the mean square of the absolute values of the differences of two identical training symbols, divided by 2. As shown in Fig.~\ref{distance}, the average SNR values for the best-case scenario at each distance are consistently higher than those of the worst-case scenario. Furthermore, the required $N_{csi}$ for the best case does not exceed 10 at any distance, whereas for the worst case, it remains below 25. Our experiments thus provide useful insights into the practical implications of selecting $N_{csi}$ in typical use-case scenarios.

\subsection{Fingerprint Matcher}
The fingerprint matcher is the process of matching the pending authenticated device with the legitimate device by comparing their fingerprints, whereas the fingerprints of legitimate devices are pre-collected and stored in a library. The result of the fingerprint matcher is then used in the decision-making process to determine whether the pending authenticated device is legitimate or not. Our fingerprint is not tailored for a specific anomaly detection algorithm. Hence, in this study, we undertake a thorough evaluation to compare the efficacy of various state-of-the-art algorithms for rogue device detection. The selected algorithms, including Isolation Forests (IForests), Local Outlier Factor (LOF), One-Class SVM (OCSVM), Density-Based Spatial Clustering of Applications with Noise (DBSCAN), and K-Nearest Neighbors (KNN), are widely recognized and used in the field of anomaly detection. In the following case study, rogue device detection is carried out using the KNN algorithm, which outperforms the other four algorithms in speed and effectiveness. See Section \ref{compare} for further information. %The fundamental principle of KNN is that the observations of the same class are expected to be in proximity to each other, and anomalies should be further away from the cluster of similar observations. The KNN algorithm detects anomalies by leveraging the relationship between data points of neighborhoods. The further a data point is from its neighbors, the more likely it is an anomaly. 

In the KNN-based fingerprint matcher, we aim to authenticate the device by comparing its denoised and normalized fingerprint $\tilde{\bold{f}}^a$ with legitimate fingerprints stored in the library. We first need to use the fingerprint $\tilde{\bold{f}}^a$ to calculate the authentication distance. Specifically, we calculate the Manhattan distances \cite{aggarwal2001surprising} between the newly arrived fingerprints $\tilde{\bold{f}}^a$ with the corresponding library fingerprints of the claimed identity. The authentication performance by using different distance metrics is compared in Section \ref{compare}. In a group of attained distances for the newly arrived fingerprint, we only keep the first \textsl{k} minimal values. We then use the average of these \textsl{k} distances as the authentication distance. Finally, a threshold is employed to authenticate the device: if the authentication distance exceeds the predetermined threshold, the device is identified as a rogue, and authentication is unsuccessful. Conversely, the device is deemed legitimate if the authentication distance falls below the threshold.

\section{A Robotic Use Case} \label{s5}
Autonomous mobile robots (AMRs) are increasingly being used to replace conveyors in production facilities due to their greater flexibility and ease of installation or replacement \cite{schneider2015design}. However, these robots also pose higher safety risks compared to conveyors since they move freely and require more extensive security measures to ensure safe operation. Securing restricted areas is a major concern for sensitive areas such as hazardous material storage areas, research labs, and equipment rooms. These areas often contain valuable or sensitive information, equipment, or materials that must be protected from unauthorized access or tampering. 
Many security measures used to grant access permissions rely on biometric information that humans possess, such as fingerprints or facial recognition. However, when it comes to robots, it can be difficult to identify them using traditional biometric measures since robots are non-biological and often look similar to each other. Therefore, it is important to develop alternative methods for establishing and verifying robot identity \cite{gavrilova2010state,yampolskiy2012artimetrics}. 
As AMRs often rely on their WiFi interfaces to communicate with other entities (e.g., area access controllers), we propose to use micro-CSI-based RF fingerprints extracted from their WiFi modules to establish robot identity, which can be used to augment conventional password-based access control. To that end, it is important to evaluate the impact of mobility on the construction of micro-CSI-based RF fingerprints, as the CSI measurements might need to be collected when robots are in motion. 

 \begin{figure}%[h]  
    \centering
\includegraphics[width=\linewidth]{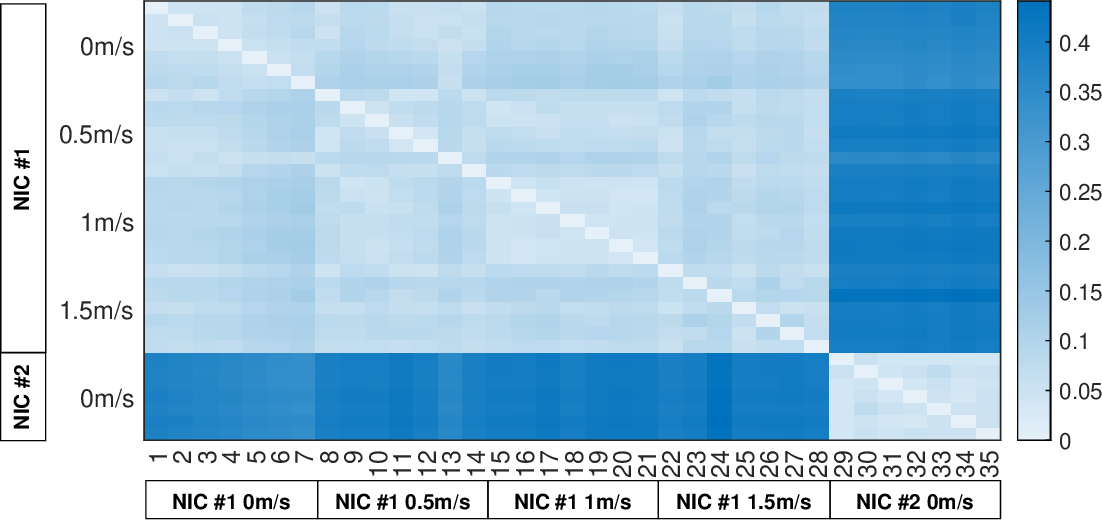}
  \caption{Mobility influence on fingerprints, where $N_{csi}=100, N_{rx}=1$.}
  \label{mobility}
  \vspace{-1em}
\end{figure}

\vspace{-1em}
\subsection{Mobility Study}
% \textcolor{red}{potential challenge} 

The movement of mobile robots can introduce Doppler shifts, which need to be taken into account when considering mobile conditions. These Doppler shifts have the potential to impact the construction of micro-CSI-based RF fingerprints. %Luckily, we find that mobility causes a negligible impact on fingerprint extraction through the following experiments.

To evaluate the impact of robot mobility on the extracted micro-CSI, we conducted an experiment using a mobile robot (AgileX Scout Mini) equipped with a mini-PC that installs an Atheros AR9271 (NIC \#1). We used the ACK CSI tool presented in Section IV-B to collect CSI. The robot was moved from a distance of 3m to 2m away from a door while collecting CSI measurements, and we repeated this process four times, each time at a distinct movement speed: 0, 0.5, 1, and 1.5 m/s. For each moving speed, we obtained 7 fingerprints from NIC \#1, where $N_{csi}=100, N_{rx}=1$. To visualize the impact of mobility on the fingerprints, we compare the intra-NIC fingerprint distances across varying speeds with inter-NIC distances between two different NICs of the same model. Fingerprints of another AR9271 (NIC \#2) were used for comparison. Note that fingerprints from NICs of the same model exhibit the highest similarity with minimal distance differentials. This similarity sets a threshold for the least distinguishability in our tests. In Fig. 8, the Manhattan distance between each pair of fingerprints was calculated and presented. We can observe that distances between fingerprints from the same NIC do not demonstrate a clear increasing trend as the moving speed of the robot increases. Besides, the distance between fingerprints from the same NIC varies in a small range and is smaller than that between different NICs of the same model. This indicates that fingerprint construction is immune to mobility and that micro-CSI-based fingerprints can be used to establish robot identity even under robot motions.

%To analyze the impact of mobility on the fingerprints, a comparison was made with the fingerprints of another NIC of the same model. Specifically, the Manhattan distance between each pair of fingerprints was calculated, and the results are presented in Fig. \ref{mobility}. We can observe from Fig. \ref{mobility} that distances between fingerprints do not demonstrate a clear increasing trend as the moving speed of the robot increases. Besides, the distance between fingerprints varies in a small range and is smaller than the distances between different NICs of the same model. This indicates that fingerprint construction is immune to mobility and that micro-CSI-based fingerprints can be used for establishing robot identity even under robot motions. % in mobile robot systems.

The key reason for our finding lies in the channel's coherence time, which greatly exceeds the transmission time of training symbols that are used for CSI estimation, rendering the Doppler-induced shift inconsequential. Generally, the maximum speed of indoor mobile robots is $v = 1.5 m/s$ \cite{speed}. Given that the frequency shift caused by the Doppler effect is $f_d = \frac{v f_c}{c}$, where $f_c$ is the carrier frequency, $c$ is the speed of waves, we can estimate that the Doppler shift in a 2.4 GHz frequency with a 20 MHz bandwidth is around 12 Hz. Assuming that the robot moves at a constant velocity, and considering the length of the training sequence (LTF) with CP ($N_l$) to be 160 and the sampling time $T_s = 50 \times 10^{-9}$ s, the time duration of the LTF sequence is $N_l T_s$. Since the time duration of the LTF sequence is much smaller than the coherence time, $N_l T_s << \frac{1}{f_d}$, we can safely assume the channel's properties remain constant during the time duration of the LTF sequence \cite{rappaport2002wireless}. Furthermore, it has been shown that the power of inter-channel interference (ICI) caused by Doppler spread has an upper bound of $P_{ICI} \leq \frac{1}{12} (2\pi f_d T_s)^2$, as demonstrated in \cite{li2001bounds}. This suggests that, in scenarios with low mobility where $f_d T_s << 1$, ICI caused by mobility can be safely disregarded. Therefore, in this context, we can deduce that the channel is invariant across the duration of the LTF sequence in general indoor mobile conditions. Therefore, the Doppler shift has a negligible impact on fingerprint extraction.

%Besides, the frequency synchronization process compensates for the frequency shift that may induced by the Doppler effect before the CSI estimation process. Therefore, the Doppler shift has a negligible impact on fingerprint extraction.}

 \begin{figure*}%[h]  
    \centering
    \begin{minipage}[]{0.6\textwidth}
      \includegraphics[width=\linewidth]{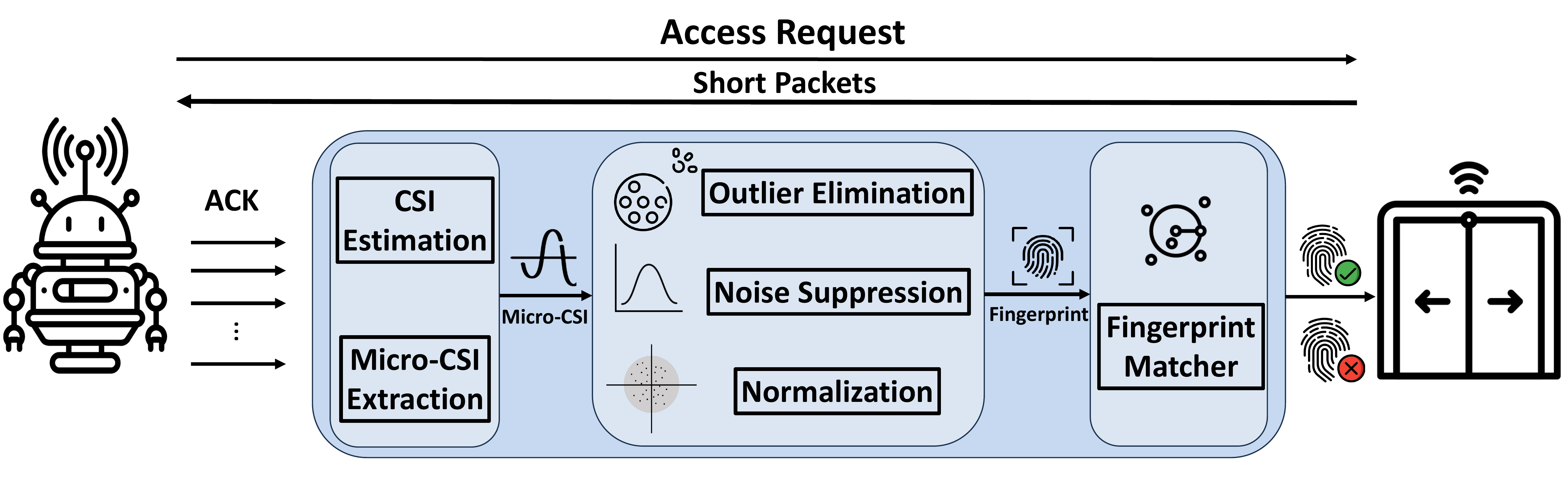}
  \caption{Robot authentication for area access control.}
  \label{system}  
    \end{minipage}
\hfill
\begin{minipage}[]{0.36\textwidth}
  % \subfloat[AgileX Scout Mini.]{\includegraphics[width=0.35\linewidth]{figures/scout_mini.jpg}}
  % \hfill
  % \subfloat[SDR for CSI collection.]{\includegraphics[width=0.35\linewidth]{figures/fpga.jpg}}
  \includegraphics[width=\linewidth]{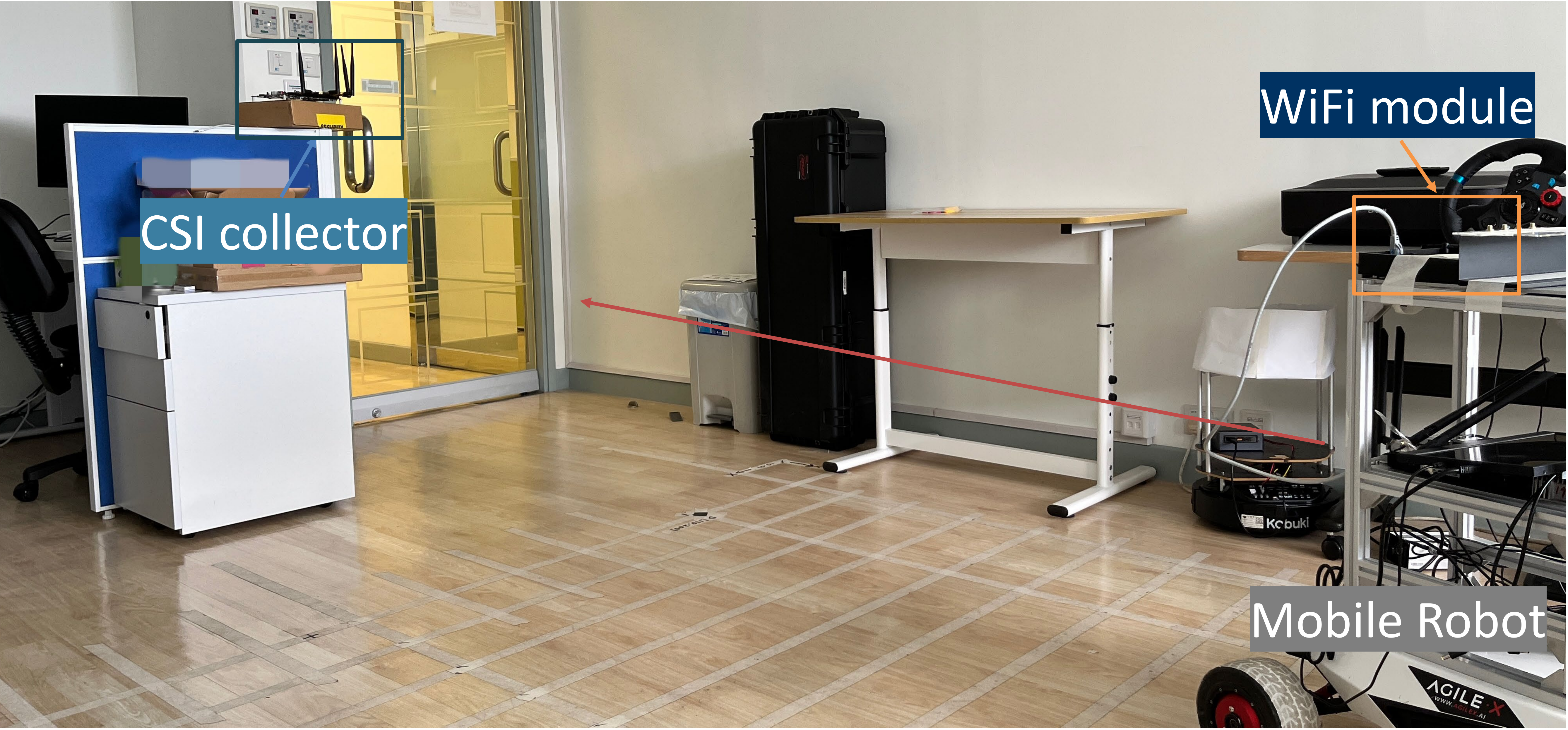}
  \caption[]{Experiment setup in room B.}
  \label{setup}
    \end{minipage}
  \vspace{-1em}
\end{figure*}

% \vspace{-1em}
\subsection{Implementation Discussions}
As we discussed in Section \ref{s4}, ACK packets are the most favorable for CSI acquisition, and a group of CSI measurements is needed to suppress noise to realize accurate authentication. To fulfill these requirements, we can implement CSI-RFF in a request-response manner. In our implementation, a robot seeking access to restricted areas is required to initiate an access request packet to the access controller situated at the autonomous door. In response, the access controller can dispatch a cluster of short packets (e.g., \textit{ping} packets) to induce the robot to reply a set of ACK packets. This exchange can be accomplished within a short time span. By adopting this approach, we can effectively enhance the authentication process for area access control, ensuring that only authorized robots are granted access to restricted areas. 

The authentication implementation of CSI-RFF for area access control in mobile robot systems is depicted in Fig. \ref{system}. CSI-RFF needs to establish the legitimate robots’ fingerprint library during the robot enrollment stage. Once a robot under authentication replies a group of ACK packets for CSI colloection, CSI-RFF first extracts micro-CSIs from CSI measurements by applying Eq. (\ref{ls}). CSI-RFF then eliminates outliers, suppresses noise, and normalizes the fingerprint by applying Algorithm \ref{fc}. Finally, the fingerprint matcher in CSI-RFF decides whether the robot is legitimate or not based on its fingerprints. If the robot passes the authentication, the door will open for legitimate access.
%Finally, CSI-RFF adopts the KNN algorithm with Manhattan distance as the criterion to measure the authentication distance. If the distance is smaller than a certain threshold, the robot passes the authentication, and the door will open for legitimate access.

% \vspace{-1em}
\subsection{Experiment Setup}
\textbf{Device Configurations}: We used 15 WiFi NICs from three manufacturers for this performance evaluation experiment, as shown in Table \ref{devices}. We set up a set of SDR devices with four antennas, $N_{rx} =4$, to collect CSI measurements as discussed in Section \ref{tool}. The network was configured to work on the 2.4GHz Channel 10 with a bandwidth of 20 MHz (i.e., the center frequency is 2457 MHz) and run the 802.11n protocol\footnote{The WiFi 5 and 6 devices used in the experiments are dual-band, capable of operating on both 2.4GHz and 5GHz bands. For our experiments, these devices were configured to operate on the 2.4GHz band to ensure that they work on the same channel as WiFi 4 devices.}. 
Every NIC that was under authentication was installed on a mini-PC carried by the robot to respond with ACK, where only one device was under authentication at any time. The experiment details are shown in Fig.~\ref{setup}. We remark that we use SDR for CSI collection because all CSI tools are not capable of reporting CSI from ACK packets for now.
% \vspace{-1em}
% \begin{table}%[h]
%   \caption{WiFi Devices}
%   \label{devices}
%   \centering
%   \resizebox{0.9\linewidth}{!}{
%   \begin{tabular}{ccccc}
%     \hline
%     NIC No.&WiFi&Brand&Model&Quantity\\
%     \hline
%     C1-C5&WiFi 4 & Espressif & ESP32-S3 & 5\\
%     C6-C9&WiFi 4 &Atheros& AR9271& 4\\
%     C10 &WiFi 5& Realtek& RTL8812BU& 1\\
%     C11 &WiFi 5& Intel& AC7260& 1\\
%     C12 &WiFi 5& Intel& AC7265& 1\\
%     C13 &WiFi 5& Intel& AC8260& 1\\
%     C14-C15&WiFi 6 & Intel& AX200 & 2\\
%   \hline
% \end{tabular}
% }
% \vspace{-1em}
% \end{table}

% \begin{figure}%[h]
%   \centering
%   \subfloat[AgileX Scout Mini.]{\includegraphics[width=0.5\linewidth]{figures/scout_mini.jpg}}
%   \subfloat[SDR for CSI collection.]{\includegraphics[width=0.5\linewidth]{figures/fpga.jpg}}
%   \caption[]{Experiment Setup.}
%   \label{setup}
%   \vspace{-1em}
% \end{figure}

\textbf{Scenarios}: We implemented our system in two different indoor scenarios in front of doors. The laboratory settings were typical network-heavy environments, potentially including cross-technology interference from Bluetooth or ZigBee.
\begin{itemize}
    \item Room A is a $10m \times 8m$ research office, which is complex with many obstacles around. The clapboards around each desk construct a complex multipath environment. %We collected CSI during busy office hours with \textit{frequent human movements}, and the transceivers were placed several meters away from each other in a line-of-sight condition.
    \item Room B is a $6m \times 5m$ common room. The common room accommodates a few desks, robots, and cabinets. %A person carried unauthorized device randomly walked within the room and in Line-of-Sight to CSI collector.
\end{itemize}

\textbf{Data Collection}: We collected CSI measurements for each NIC in Room A and Room B during busy office hours, respectively. Collections in Room A are used to establish the fingerprint library, and collections in Room B are used for testing. The time gap between collections of different rooms is around 15 days to ensure the collected CSI space in time and environments for better evaluating the robustness of CSI-RFF.

In each room, we conducted two CSI collections for each NIC under static and mobile conditions, respectively. In the static condition, the robot was positioned 2 meters away from the door, and the SDR was located near the door. In the mobile condition, the robot moved from a distance of 3 meters to 2 meters away from the door at a speed of 0.5 m/s during CSI collection. The SNR for the gathered CSI data spanned from 0.67 dB to 47 dB, with the mean SNR resting at 34 dB. Considering packet losses during each CSI collection, we obtained approximately 1000 CSI measurements from ACK packets for each NIC. In total, we used approximately $1000\text{ (Packets/collection)}\times 2\text{ (Collections)}=2000$ CSI measurements collected in Room A for each NIC to establish the fingerprint library. Another dataset, encompassing around 2000 CSI measurements, was subsequently gathered in Room B for the purpose of authentication in the evaluations that followed.

\vspace{-1em}
\subsection{Open-Set Authentication Performance} 
We adopt the attack detection rate (ADR) and false alarm rate (FAR) as performance metrics, where the ADR is the probability of successfully detecting a rogue device, and the FAR is the probability of mistakenly rejecting a legitimate device. 
In all the results presented in this section, the number of fingerprints to construct the fingerprint library of each legitimate device is $S=2000/N_{csi}$, where each group of $N_{csi}$ CSI measurements constructs one fingerprint. To challenge our framework, we used fingerprints of legitimate devices collected in Room A to establish the fingerprint library and fingerprints of all devices collected in Room B as test fingerprints. %This setup evaluates the capabilities of CSI-RFF in accepting legitimate devices under different environments
This setup rigorously evaluates the robustness of CSI-RFF in real-world settings where environmental changes are commonplace.\footnote{When fingerprints from Room B are used for the library and fingerprints from Room A are used for authentication, CSI-RFF achieves similar performance.}. Besides, we set $\textsl{k}=\sqrt{S}$ in our KNN algorithm, i.e., the distance is the average Manhattan distance between each test fingerprint and its \textsl{k} nearest neighbors stored in the fingerprint library. Generally, $\textsl{k}=\sqrt{S}$ is a balancing choice to avoid overfitting or underfitting.

% \vspace{-1em}
\begin{table}%[h]
  \caption[ADR]{Attack Detection Rate. In each grid, the percentage is the average ADR performance of CSI-RFF.}
  \centering
\subfloat[Static Conditions]{
  \begin{tabular}{l|l|l|l|l}
    \hline $N_{csi}$&1&5&10&20\\
    \hline
      \multirow{1}{4em}{FAR=0\%}
      & 61.26\%&91.57\%& 97.82\% & 99.96\%\\ 

    % & 26.00\%&90.83\%& 96.94\% & 99.92\%\\
  % \hline
    %     \multirow{1}{4em}{FAR=0.1\%}
    % & 17.50\%&88.89\%& 96.89\% & 99.86\%\\
  \hline
          \multirow{1}{5em}{FAR$\leq$1\%}
          & 84.75\%&97.18\%& 98.65\% & 99.98\%\\

    % & 65.16\%&93.54\%& 98.01\% & 99.92\%\\
  \hline
\end{tabular}
\label{static}
}
\newline
% \vspace*{0.3 cm}
% \newline
\subfloat[Mobile Conditions]{
\label{mobile}
  \begin{tabular}{l|l|l|l|l}
    \hline $N_{csi}$&1&5&10&20\\
    \hline
      \multirow{1}{4em}{FAR=0\%}
      & 61.87\%&91.14\%& 97.30\% & 99.57\%\\ 

    % & 39.45\%&89.00\%& 96.29\% & 99.77\%\\
  % \hline
  %       \multirow{1}{4em}{FAR=0.1\%}
  %   & 23.72\%&87.87\%& 94.90\% & 99.63\%\\
  \hline
          \multirow{1}{5em}{FAR$\leq$1\%}
          & 84.88\%&96.80\%& 98.43\% &  99.996\%\\
    % & 69.57\%&92.87\%& 97.93\% &  99.77\%\\
  \hline
\end{tabular}
}
\label{adr}
 %\vspace{-0.5em}
\end{table}

\textbf{Static Conditions.} We first evaluate in static conditions. We made each of the 15 NICs to be the legitimate device in turn and to be attacked by the remaining 14 NICs. Fingerprints of the legitimate device collected in Room A construct the fingerprint library. All fingerprints of 15 NICs collected in room B are used for testing. 
The detailed ADR and FAR are shown in Table \ref{adr}a, where the percentage in each grid of the table is the average ADR among all devices as the legitimate device in turn in static conditions. The experimental results show that CSI-RFF is resistant to environmental changes. Furthermore, as the number of CSI measurements used for constructing one fingerprint (i.e., $N_{csi}$) increases, CSI-RFF achieves 99.96$\%$ ADR with 0$\%$ FAR including attacker and legitimate NICs are different models or the same model. 
%It is important to note that the performance achieved in Table \ref{adr}a was obtained using both static and mobile data collected in Room A to establish the fingerprint library. However, when using only static data from Room A to establish the fingerprint library and setting $N_{csi}=20$, the performance drops to 98.41$\%$ ADR with 0$\%$ FAR. This decrease in performance is due to the fact that the intensity of extracted micro-CSI-based fingerprints varies slightly among different channels, as discussed in Section \ref{robustness}. As a result, fingerprints collected under mobile conditions cover a wider range of fingerprint variants and can achieve more accurate authentication.

{\textbf{Mobile Conditions.} In our second evaluation, we evaluate CSI-RFF in mobile conditions. The evaluation setup is the same as the first evaluation.} The detailed ADR and FAR are shown in Table \ref{adr}b, where the percentage in each grid is the average ADR among all devices in mobile conditions.
We observed that the performance of our method under mobile conditions is comparable to the performance achieved when devices remain stationary. Specifically, CSI-RFF achieves a 99.57$\%$ ADR with 0$\%$ FAR when $N_{csi} =20$. The experimental results presented in Table \ref{adr}b demonstrate that micro-CSI-based fingerprint is highly resistant to device mobility. 
%It should be noted that the slight variance in performance between static and mobile conditions can be attributed to the occurrence of abnormal cases of collected CSI. When $N_{csi}$ increases, our algorithm exhibits an improved ability to eliminate these abnormal cases. Consequently, both static and mobile data can achieve similar performance.

% \vspace{-1em}
\subsection{Discussions}
\label{compare}
We now discuss the impact of different factors in CSI-RFF.

\begin{figure*}%[h]
  \centering
  \begin{minipage}[]{0.48\textwidth}
 \includegraphics[width=\linewidth]{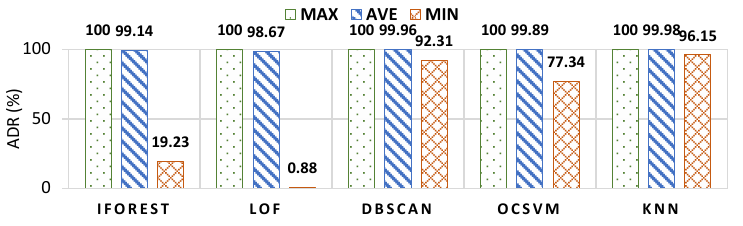}
  \caption{Impacts of anomaly detection algorithms, when FAR $\leq 1\%$ and $N_{csi}=20, N_{rx}=4$.}
  \label{comp_al}
  \vspace{0.5em}
  \end{minipage}
\hfill
    \begin{minipage}[]{0.48\textwidth}
  \includegraphics[width=\linewidth]{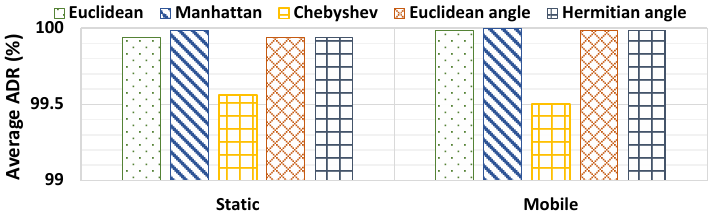}
  \caption{Impacts of distance metrics, when FAR $\leq 1\%$ and $N_{csi}=20, N_{rx}=4$. }
  \label{impact}
  \end{minipage}

    \begin{minipage}[]{0.48\textwidth}
  \includegraphics[width=\linewidth]{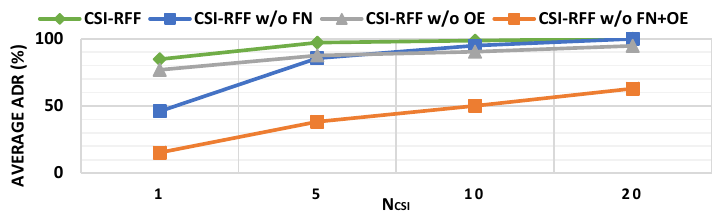}
  \caption{Impacts of fingerprint normalization (FN) and outlier elimination (OE) under static conditions, where FAR $\leq 1\%$ and $N_{csi}=20, N_{rx}=4$..}
  \label{ablation}
  % \vspace{-1em}
  \end{minipage}
 \hfill 
    \begin{minipage}[]{0.48\textwidth}
  \includegraphics[width=\linewidth]{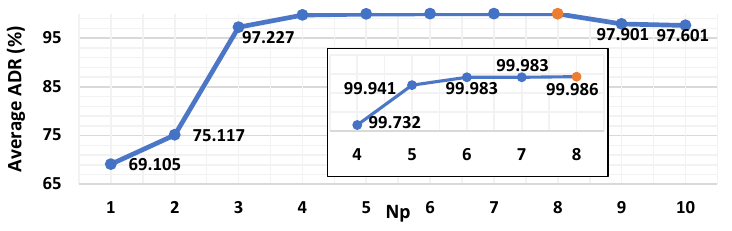}
  \caption{Impacts of $N_p$, when FAR$\leq 1\%$ and $N_{csi}=20, N_{rx}=4$..}
  \label{np}
  \end{minipage}
  \vspace{-1em}
\end{figure*}

\textbf{Anomaly Detection Algorithms.} To compare the performance of different anomaly detection algorithms, we tested the IForest, LOF, DBSCAN, OCSVM, and KNN algorithms. The performance metrics evaluated include maximum (MAX), average (AVE), and minimum (MIN) values of ADR when FAR $\leq1\%$, and $N_{csi}=20$. From Fig.~\ref{comp_al}, we can observe that all algorithms achieved high maximum performance scores of 100$\%$. We noticed that such a score was achieved when legitimate NIC and attacker were of different models. This indicates that these algorithms are effective and comparable in detecting anomalies when NICs are of different models. 
%In terms of average performance, the Iforest algorithm achieved the lowest score of 99.14$\%$ and KNN exhibited the highest average performance score of 99.98$\%$. LOF, OCSVM, and KNN demonstrated similar average performance scores of 99.82$\%$, 99.85$\%$, and 99.87$\%$, respectively. 
Examining the minimum performance scores, we observed that LOF and IForest achieved extremely low scores (0.88$\%$ and 19.23$\%$, respectively) compared to the other algorithms. DBSCAN and OCSVM demonstrated a performance score of 92.31$\%$ and 77.34$\%$, respectively. We noticed that these minimum scores were mainly contributed by the cases when legitimate NIC and attacker are of the same model. On the other hand, KNN exhibited the highest minimum performance scores of 96.15$\%$, suggesting their robustness in detecting anomalies even in challenging scenarios. Based on the analysis, it can be concluded that all algorithms performed well overall, with around 99$\%$ averaged ADR. However, KNN is superior to other algorithms in challenging scenarios. Besides, KNN is the fastest algorithm according to our implementation. %\red{When compared with IForest, KNN excels with high-dimensional CSI data due to its reliance on the local proximity of data points. Unlike LOF, DBSCAN, and OCSVM, which operate under certain assumptions about the data distribution, the non-parametric nature of KNN allows it to effectively navigate the data space without such constraints. Consequently, KNN performs better in complex-valued high-dimensional CSI data.}

Upon further analysis, we can attribute the different performances of these algorithms to their intrinsic methodology. The random partitioning strategy of IForest may miss subtle structures within complex-valued fingerprints, and LOF's performance can suffer from environmental changes affecting local density in training and testing fingerprints. DBSCAN, however, compensates by considering global density, which improves robustness. OCSVM is effective in distinguishing outliers by using kernel transformation but is noise-sensitive. KNN, as a non-parametric method, demonstrates superior performance by adapting to the data space without assuming any underlying data distribution, which explains its enhanced effectiveness in our evaluation.

\textbf{Distance Metrics.} The choice of distance metric in the KNN algorithm may affect the authentication performance. In our experiment, we compared the authentication performance when using various popular distance metrics for complex vectors in the KNN algorithm, including Chebyshev distance, Euclidean distance, Manhattan distance, Hermitian angle, and Euclidean angle \cite{scharnhorst2001angles}. Fig. \ref{impact} plots the average ADR of 15 NICs as the legitimate device in turn and be attacked by the rest of NICs when FAR$\leq1\%$. We can observe that Manhattan distance outperforms other metrics in both static and mobile conditions. The Chebyshev metric, however, exhibited a relatively lower accuracy because it does not consider the actual distances between the fingerprints but focuses on the largest difference. Overall, our experimental results reveal that the Manhattan distance is the most suitable one in our CSI-RFF framework. We also note that the superior performance of Manhattan distance for high-dimensional data has also been demonstrated in \cite{aggarwal2001surprising}.

\textbf{Fingerprint Normalization and Outlier Elimination.} We now discuss the impact of the fingerprint normalization (FN) and outlier elimination (OE) process in our framework, as described in Algorithm \ref{fc}. These strategies were specifically designed to address the challenges posed by variations in fingerprints and the presence of outliers. To do so, we compared the averaged ADR with and without FN and/or OE while maintaining FAR$\leq1\%$. The results under static conditions are illustrated in Fig. \ref{ablation}, and the results under mobile conditions yielded similar results, which are omitted for brevity. The results clearly demonstrate the effectiveness of incorporating the FN strategy. By normalizing the fingerprints, CSI-RFF achieved significantly higher accuracy across all $N_{csi}$ values, especially when $N_{csi}$ is small. More specifically, the improvements are 38.75$\%$, 11.71$\%$, 3.76$\%$, 0.03$\%$,  when $N_{csi} = 1,~5,~10,~20$, respectively. Besides, comparing performance without and with outlier elimination, the ADR drops significantly, around 10$\%$ for all $N_{csi}$ values. This drop in performance is due to the presence of outliers in the CSI data, which can largely affect the accuracy of our authentication methods. Furthermore, we examined the combined impact of both the FN and OE strategies (i.e., the CSI-RFF curve in Fig. \ref{ablation}). The results indicated that employing these strategies together further enhanced the performance, with at least a 37.08$\%$ increment compared to the scheme without them. %This highlights the synergistic effect of applying both strategies, leading to improved accuracy and reliability in authentication.

\textbf{Influence of $N_p$.} The choice of $N_p$ can affect the  proposed fingerprint extraction method (\ref{ls}). To investigate this influence, we conducted a comparison of the authentication performance for different values of $N_p$ under static conditions with $N_{csi}=20$. Fig. \ref{np} plots the average ADR of 15 NICs as the legitimate device in turn and be attacked by the rest of NICs under different $N_p$, where FAR$\leq 1\%$. We can observe that the best performance is achieved when $N_p=8$. Small values of $N_p$ lead to insufficient elimination of channel information, and large values of $N_p$ result in more hardware distortions being eliminated from the estimated fingerprints.

Furthermore, our algorithm can also accommodate CSI from various packet types using different numbers of subcarriers by adjusting $N_p$ proportionally to the channel bandwidth. For instance, in a 20MHz bandwidth, the pulse impulse is presumed to decay to zero after 8 samples, and this number of samples with the decay interval scales with bandwidth; doubling the bandwidth leads to halving the sample interval and, consequently, doubling $N_p$. It is noteworthy that our CSI measurements are sourced from the physical-layer ACK packets, which, according to our experiments, use legacy format (802.11a/g) with 52 subcarriers, irrespective of operational bandwidth specified by WiFi protocols. This uniformity ensures the consistency of both $N_p$ and the dimension of fingerprints, thereby facilitating the algorithmic implementation.
%  \begin{figure}%[h]  
%     \centering
%   \includegraphics[width=\linewidth]{figures/N_p.eps}
%   \caption{Impacts of $N_p$, when FAR$\leq 1\%$ and $N_{csi}=20$.}
%   \label{np}
%   \vspace{-1em}
% \end{figure}

%  \begin{figure}%[h]  
%     \centering
%   \includegraphics[width=\linewidth]{figures/distances_compare.pdf}
%   \caption{Impacts of distance metrics, when FAR $\leq 1\%$.\textcolor{red}{why} }
%   \label{impact}
%   \vspace{-2em}
% \end{figure}

\textbf{Influence of Bandwidth and Carrier Frequency}. In real-world applications, network configurations may occasionally change due to various factors such as strong interference or network management policies. When the bandwidth changes, the pulse shaping filter is adjusted accordingly to match the operational bandwidth. Such adjustments could potentially change the fingerprint characteristics. Moreover, as the oscillator is part of the hardware RF chain, variations in carrier frequency may also affect fingerprints. Our empirical findings indicate that the extent of these influences varies among different NICs. Nevertheless, such variations do not compromise the validity of our CSI-RFF framework. This is because our experimental results show that the fingerprints obtained at each bandwidth and carrier frequency are consistently stable and distinguishable between devices. 
We conducted more experiments on WiFi 2.4GHz channel 1 and 5GHz channel 161 to further substantiate our claims. For the experiment on channel 1, we used fingerprints from 15 NICs, all collected in two different LoS positions within the same room, to create our fingerprint library and carry out the authentication process. In the 5GHz experiment, due to the incompatibility of WiFi 4 devices with the 5GHz band, we used the rest 6 WiFi 5 and 6 NICs to perform authentication. Results from both experiments demonstrated a 100\% ADR with a 0\% FAR when $N_{csi}=20, N_{rx}=4$. It is noteworthy that the improved performance relative to results in Table II obtained on channel 10 can be attributed to the consistent environmental conditions provided by constructing the fingerprint library and conducting the authentication within the same room, which minimizes environmental variability. Therefore, our system is adaptive and can recalibrate to different channels and bandwidths, ensuring the reliability of the authentication. 
%all ack use 20mhz, 52 subcarrier; fingerprint may vary with channel, depend on manufacture accuracy. so we conduct experiment to verify the distinguishability.  we assume ap will not change channel, and it is easy for ap to get fingerprints of legitimate devices. experiments

\textbf{Overhead.} Note that a group of CSI measurements (i.e., $N_{csi}$) are needed to suppress noise and outliers for realizing accurate authentication. We remark that even when $N_{csi}=20$, if devices under authentication reply to the challenges with ACK packets and emit one packet approximately every 50 microseconds as configured in our experiment, the time consumption for CSI collection is just around 1 millisecond, and the overhead of CSI collection is $27~\text{(bytes/packets)} \times 20 ~\text{(packets)} \div 1 ~\text{(millisecond)} = 540~\text{kbps}$, which will not introduce a noticeable impact on the network throughput. Besides, Table \ref{comparision} compares CSI-RFF with the two most relevant related works that extract RF fingerprints from CSI for authentication purposes. The second column represents the number of devices used in their evaluation setup. According to Table \ref{comparision}, we can conclude that CSI-RFF achieves better performance by using much fewer CSI measurements in both static and mobile conditions. Specifically, compared to the scheme in \cite{hua}, CSI-RFF can achieve $10\times$ less FAR by using $250$ times less CSI measurements while obtaining a slightly higher ADR. Besides, the CFO fingerprint used in \cite{hua} may not work in mobile scenarios. Furthermore, compared to the scheme in \cite{lin2020}, CSI-RFF can decrease the FAR by over $30$ times and the number of required CSI measurements by $100$ times while increasing the ADR.
\begin{table}%[h]
    \caption{Comparison of CSI-RFF with related works. $\#$ in the second column represents the number of devices used. For a fair comparison, {the performance of the scheme in \cite{lin2020} was achieved using CSI-based fingerprint only.}}
    \centering
    \resizebox{\linewidth}{!}{
    \begin{tabular}{l|l|l|l}
    \hline
     Fingerprint & $\#$ &Performance& $N_{csi}$ \\
    \hline
    CFO \cite{hua}&8 & ADR=97.24\%, FAR=1.47\%& 5000\\
    %\hline
   % I/Q\cite{liu}&3/30&&200\\
    \hline
    PA \cite{lin2020}&12&ADR=93\%, FAR=3\%&2000\\
    \hline
    Micro-CSI&15&ADR=99\%, FAR=0\%&20\\
    \hline
    \end{tabular}
    }
    \label{comparision}
    \vspace{-1.5em}
\end{table}

\section{Limitations and Future Work}
When the number of devices of the same model increases, solely depending on a single fingerprint (feature) to realize accurate authentication becomes less practical due to higher fingerprint similarity. In future work, we will evaluate and improve the performance of CSI-RFF in large-scale systems by collecting the CSI of more WiFi devices of the same model and diverse brands. We plan to investigate the following aspects. First, combining different features can effectively decrease the fingerprint similarity between devices and tolerate more channel noise and processing errors. To that end, we will combine other features extracted from CSI measurements and explore the best feature combination scheme. For example, location signatures extracted from CSI measurements (e.g., angle of arrival) can be combined with micro-CSI-based fingerprints to boost authentication performance. Second, we will try out more advanced authentication algorithms (e.g., deep learning-based approaches), which are expected to achieve better performance than the considered KNN algorithm.

The efficacy of our approach is primarily demonstrated in LoS scenarios, with the performance expected to decline under non-line-of-sight (NLoS) conditions due to multipath interference, leading to fewer distinct, unoccupied channel taps for micro-CSI extraction. Addressing this, we recognize the need for advanced techniques that can better differentiate between channel effects and hardware distortions in such challenging environments. Future work will explore more advanced approaches (e.g., deep learning-based approaches) to disentangle channels from hardware distortions. Such approaches may include the development of neural network architectures and training strategies that are thoughtfully designed for micro-CSI to eliminate multipath interference, thereby achieving robust micro-CSI-based fingerprinting in NLoS scenarios.

% \vspace{-1em}
\section{Conclusions}
In this paper, we developed a new RF fingerprinting-based open-set authentication framework, termed CSI-RFF, which leverages micro-signals on channel state information (CSI) curves to construct fingerprints for authenticating commercial WiFi devices. We conducted experiments to confirm that the micro-signals on CSI (i.e., micro-CSI) most likely originate from RF circuitry imperfections and to unveil that micro-CSI exists on signals transmitted by WiFi 4/5/6 NICs. We developed a signal space-based extraction algorithm that effectively disentangles distortions caused by wireless channels and hardware imperfections under line-of-sight (LoS) scenarios. We systematically evaluated the device uniqueness, time invariance, and mobility independence of micro-CSI-based fingerprints based on CSI measurements collected across 14 months. Finally, we demonstrate the application of CSI-RFF to identify robots based on the signals transmitted by their WiFi interfaces in an area access control use case. Specifically, we implement CSI-RFF to authenticate 15 commercial off-the-shelf (COTS) WiFi 4/5/6 NICs carried by a mobile robot. Our experimental results showed that CSI-RFF can achieve close to $99 \%$ attack detection rate (ADR) with 0$\%$ false alarming rate (FAR) in both static and mobile conditions with training and testing data collected in different rooms. When compared with the latest CSI-based RF fingerprinting schemes in the literature, CSI-RFF can decrease the FAR by more than one order of magnitude while using fewer CSI measurements and achieving slightly higher ADR. 

% \vspace{-1em}
 \bibliographystyle{IEEEtran}
 \bibliography{ref} 

% \begin{IEEEbiography}
% [{\includegraphics[width=1in,height=1.25in,clip,keepaspectratio]{Ruiqi_photo_4v5.jpg}}]{Ruiqi Kong} (Student Member, IEEE) received the B.S. degree in Communication and Information Engineering from the University of Electronic Science and Technology of China, in 2020. She is currently working towards a Ph.D. degree at the Department of Information Engineering, The Chinese University of Hong Kong, SAR, China. Her research interests include wireless networking and security with an emphasis on the physical layer.
% \end{IEEEbiography}
% \vspace{-14pt}

% \begin{IEEEbiography}
% [{\includegraphics[width=1in,height=1.25in,clip,keepaspectratio]{}}]{He (Henry) Chen} (Member, IEEE) received the Ph.D. degree in electrical engineering from The University of Sydney, Sydney, Australia, in 2015. He was a Research Fellow with the School of Electrical and Information Engineering, The University of Sydney. In July 2019, he joined the Department of Information Engineering, The Chinese University of Hong Kong, as a Faculty Member, where he is currently an Assistant Professor. His current research interests include low-latency wireless communications and networking, and their applications in robotic systems. From 2020 to 2022, he served on the editorial board for IEEE WIRELESS COMMUNICATIONS LETTERS. He is serving on the editorial board for IEEE TRANSACTIONS ON WIRELESS COMMUNICATIONS.
% \end{IEEEbiography}

\end{document}